\documentclass[journal,letterpaper,twocolumn,final,10pt]{IEEEtran}
\usepackage{amsthm}
\usepackage{amsmath}
\usepackage{amsfonts}
\usepackage{dsfont}
\usepackage{mathrsfs}
\usepackage{pifont}
\usepackage{amssymb}
\usepackage{verbatim}
\usepackage{upgreek}
\usepackage{color}
\usepackage[ruled]{algorithm2e} 
\usepackage[final]{graphicx}
\usepackage{lipsum} 
\usepackage{titlesec}
\usepackage{epsfig}
\usepackage{eucal}
\usepackage{cite}
\usepackage{subfigure}
\usepackage{cuted}
\usepackage{caption}
\usepackage{booktabs}

\newcommand{\bbm}{\begin{bmatrix}}
\newcommand{\ebm}{\end{bmatrix}}

\newcommand{\bit}{\begin{itemize}}
\newcommand{\eit}{\end{itemize}}

\newcommand{\ben}{\begin{enumerate}}
\newcommand{\een}{\end{enumerate}}

\newcommand{\bdesc}{\begin{description}}
\newcommand{\edesc}{\end{description}}

\newcommand{\bea}{\begin{array}}
\newcommand{\eea}{\end{array}}

\newcommand{\tr}{\mbox{\rm Tr}\, }

\newcommand{\beqa}{\begin{eqnarray}}
\newcommand{\eeqa}{\end{eqnarray}}

\newcommand{\Comment}[1]{}

\def\C{{\mathds C}}

\def\cA{\mbox{$\mathcal A$}}

\def\cC{\mbox{$\CMcal C$}}

\def\cN{\mbox{$\CMcal N$}}

\newcommand{\be}{\begin{equation}}
\newcommand{\ee}{\end{equation}}

\newcommand{\bzero}{{\mbox{\boldmath $0$}}}
\def\cX{\mbox{$\mathcal X$}}
\def\cY{\mbox{$\mathcal Y$}}

\newcommand{\bn}{{\mbox{\boldmath $n$}}}
\newcommand{\bm}{{\mbox{\boldmath $m$}}}

\newcommand{\bv}{{\mbox{\boldmath $v$}}}
\newcommand{\bx}{{\mbox{\boldmath $x$}}}

\newcommand{\bz}{{\mbox{\boldmath $z$}}}

\newcommand{\bA}{{\mbox{\boldmath $A$}}}
\newcommand{\bB}{{\mbox{\boldmath $B$}}}

\newcommand{\bH}{{\mbox{\boldmath $H$}}}
\newcommand{\bI}{{\mbox{\boldmath $I$}}}

\newcommand{\bM}{{\mbox{\boldmath $M$}}}

\newcommand{\bP}{{\mbox{\boldmath $P$}}}

\newcommand{\bR}{{\mbox{\boldmath $R$}}}
\newcommand{\bS}{{\mbox{\boldmath $S$}}}

\newcommand{\bX}{{\mbox{\boldmath $X$}}}

\newcommand{\bZ}{{\mbox{\boldmath $Z$}}}

\newcommand{\diag}{\mbox{\boldmath\bf diag}\, }

\newcommand{\balpha}{{\mbox{\boldmath $\alpha$}}}
\newcommand{\bbeta}{{\mbox{\boldmath $\beta$}}}

\newcommand{\bnu}{{\mbox{\boldmath $\nu$}}}

\newcommand{\test}{\mbox{$
\begin{array}{c}
\stackrel{ \stackrel{\textstyle H_1}{\textstyle >} }{
\stackrel{\textstyle <}{\textstyle H_0} }
\end{array}
$}}

\title{ Adaptive Radar Detection in joint Range and Azimuth based on the Hierarchical Latent Variable Model}

\author{Linjie Yan, \IEEEmembership{Member, IEEE}, Chengpeng Hao, \IEEEmembership{Senior Member, IEEE}, Sudan Han, Giuseppe Ricci, \IEEEmembership{Senior Member, IEEE}, Zhanhao Hu, and Danilo Orlando, \IEEEmembership{Senior Member, IEEE} 
	
\thanks{Linjie Yan, Chengpeng Hao and Zhanhao Hu are with the Institute of Acoustics, Chinese Academy of Sciences, Beijing 100864, China, E-mail: {\tt yanlinjie16@163.com; haochengp@mail.ioa.ac.cn; huzhanhao@mail.ioa.ac.cn}.}
\thanks{Sudan Han is with  the  National  Innovation  Institute  of  Defense  Technology, Beijing, China E-mail: xiaoxiaosu0626@163.com.}
\thanks{Giuseppe Ricci is with the Dipartimento di Ingegneria dell'Innovazione,
	Universit\`a del Salento, Via Monteroni, 73100 Lecce, Italy.
	E-mail: {\tt giuseppe.ricci@unisalento.it}.}
\thanks{Danilo Orlando is with Dipartimento di Ingegneria dell' Informazione, Universit\`a di Pisa, Via Caruso 16, 56122 Pisa, Italy. E-mail: {\tt danilo.orlando@unipi.it}. \emph{(Corresponding author: Danilo Orlando.)}}

\thanks{The work of D.Orlando was partially supported by the Italian Ministry of Education and Research (MUR) in the framework of the FoReLab project (Departments of Excellence) and in part by the European Union in the NextGenerationEU plan through the Italian program ``Bando PRIN 2022," D.D. 104/2022 (PE7, project ``CIRCE," code H53D23000420006). The work of Giuseppe Ricci was supported in part by the European Union in the NextGenerationEU plan through the Italian program ``Bando PRIN 2022,” D.D. 104/2022 (PE7, project ``CIRCE,” code F53D23000450006). The work of Linjie Yan, Chengpeng Hao, and Zhanhao Hu was supported	by the National Natural Science Foundation of China under Grant 62201564.  The work of Sudan Han was supported by the National Natural Science Foundation of China under Grant 62201607. }

}

\begin{document}

\maketitle

\begin{abstract}

This paper focuses on the design of a robust decision scheme capable of operating in target-rich scenarios with unknown signal signatures (including their range positions, angles of arrival, and number) in a background of Gaussian disturbance. To solve the problem at hand, a novel estimation procedure is conceived resorting to the  expectation-maximization algorithm in conjunction with the hierarchical latent variable model that are exploited to come up with  a maximum \textit{a posteriori} rule for  reliable signal classification and angle of arrival estimation. The estimates returned by the procedure are then used to build up an adaptive detection architecture in range and azimuth based on the likelihood ratio test with enhanced detection performance. Remarkably, it is shown that the new decision scheme can maintain  constant the false alarm rate when the interference parameters vary in the considered range of values. The performance assessment, conducted
by means of Monte Carlo simulation, highlights that the proposed detector exhibits superior  detection performance in comparison with the existing GLRT-based competitors.

\end{abstract}


\section{Introduction}
\label{Sec:Introduction}

In radar applications, the adaptive target detection in Gaussian environments faces significant challenges due to imperfections in signal direction information, commonly referred to as Angle of Arrival (AoA) mismatch. This issue can generally be compartmentalized into two main situations. One instance arises from the fact that the nominal AoA, assumed or expected direction of the target echoes, deviates from the actual AoA because of calibration and pointing errors, imperfect antenna shape, and wavefront distortions \cite{5438467, 2010Adaptive}. The other case occurs in the presence of large main beams, wherein the target AoA is completely unknown, leading to an increased uncertainty and complexity in the detection task. In the open literature on radar Space-Time Adaptive Detection (STAD), most conventional approaches, including the Generalized Likelihood
Ratio Test (GLRT) \cite{4104190}, adaptive matched filter \cite{135446}, adaptive coherence estimator \cite{599116}, etc., are typically designed assuming that the radar pointing direction coincides with the actual target direction, namely, the nominal and actual steering vectors perfectly match. It is apparent that this assumption is not often valid from an operating point of view, inevitably leading to a degradation of detection performance in the presence of AoA mismatch.

In \cite{4102349,18674,7303927}, the authors provide an extensive  analysis for the performance of STAP detectors under mismatched conditions by resorting to the generalized cosine-squared  between the actual and nominal signal direction  as a  metric to quantify the discrepancy. Over the past few decades, various robust detectors have been proposed and assessed for different operating scenarios to mitigate the target energy loss resulting from AoA mismatch. A conventional strategy for addressing angle uncertainty involves the use of subspace-based model \cite{301849,381913,492544,890324}. Such detectors assume that target signature belongs to a subspace spanned by specific vectors representative of the angular sector of interest. For instance, the subspace can be obtained by computing the eigenvectors corresponding to the most dominant eigenvalues of the slepian matrix \cite{VanTrees2002}. Thus the known rank-one signal model is replaced by an unknown linear combination of the  base vectors whose   coefficients are suitably estimated.  For instance, in \cite{381913}, an adaptive scheme is introduced for detecting a signal modeled as an  unknown deterministic vector with unspecified correlation properties in the Homogeneous Environment (HE), namely when data under test  and training samples are affected by interference with the same statistical properties. To account for AoA uncertainties in Partially HE (PHE) wherein the interference components of primary and secondary data have different power levels, a GLRT-based detector is derived in \cite{4014367} based on the framework developed in \cite{1542473}. Furthermore, \cite{GLRT-based,1145714}, and \cite{6576130} present several subspace-based detectors obtained through different design criteria to address the challenge of detecting range-distributed targets in case of AoA mismatch. For more details, the interested reader is referred to \cite{1396421,4531367,5978230,6879481,7938714}. An alternative approach to enhance robustness under
uncertain AoAs leverages the cone idea,
which constrains the signal signature within a cone centered
on its nominal value, as applied in \cite{ 1561887, 4203039, 4895239, 8716559}  for both point-like
and distributed targets. Finally, it is worth mentioning the two-stage detectors obtained by cascading two detectors to achieve a flexible trade-off between robustness and rejection of unwanted signals \cite{839972,4531361,5165249, 2011Performance}.

However, the detectors discussed so far do not return or exploit any (even though rough) estimate of the actual AoA, possibly leading to a consequent loss in  detection performance.  In light of this problem, some detection architectures with AoA estimation capabilities have been proposed. Such solutions are based on compressed sensing  for scenarios where the attack of Coherent Jammers (CJs) and/or Noise-Like Jammers (NLJs) can occur. The interested reader is referred to \cite{8781902} for an example of a novel detection stage that incorporates the so-called sparse learning via iterative minimization algorithm \cite{5617289} and returns rough estimates of AoA for the discrimination between the target and CJs. Similarly, the method proposed in \cite{9076078} is capable of detecting multiple NLJs while offering approximate AoA estimates. In \cite{9040449}, 
the sparse reconstruction technique is employed in the design of tunable adaptive architectures with an enhanced selectivity obtained through the assistance of AoA estimation. It is important to note that most of these existing detection architectures are designed for either a point-like target located in a single cell under test or a range-spread target whose echoes are  scattered across contiguous range bins. However, nowadays, the spread of unmanned/autonomous vehicles in addition to common moving objects, have made it necessary for modern radar to operate in target-rich scenarios. As a consequence, radar community focused on the development of architectures  capable of detecting multiple point-like targets. In \cite{1605248} and \cite{10058041}, some related examples are provided. Specifically, the  detection schemes  in  \cite{1605248} are designed in HE to detect extended and multiple point-like targets without assignment of secondary data. In  \cite{10058041}, a more adverse environment is considered in the presence of heterogeneous clutter edges as well as multiple targets.  In this context, several GLRT-based detectors are devised  resorting to the Expectation-Maximization (EM) algorithm in conjunction with cyclic optimization procedures to jointly address the challenges of targets detection and clutter classification.
Again, notice that these methods are designed under the assumption that the multiple target echoes originate from the same known nominal direction, and, hence, they do not  capitalize all the available energy in case of mismatch. To overcome this limitation, a detector performing AoA estimation  would achieve better detection performance in target-dense situations with unspecified AoAs. 

 With the above remarks in mind, in this paper, we make a further move in the context of adaptive radar  detection in  Gaussian background with target signature mismatch. Specifically, we explore a more challenging scenario that takes into account both the presence of multiple point-like targets and the uncertainty related to their different  AoAs as well as their number. Addressing this lack of information, that is exploited by the conventional detectors, through the maximum likelihood approach leads to optimization problems that are intractable from the point of view of mathematics and computational requirements. For this reason, we resort to a hierarchical Latent Variable Model (LVM)  \cite{9048459, 10465106} by introducing {\em outer} hidden random variables that allow us to establish the range bins occupied by targets and  {\em inner} hidden random variables to handle the uncertainty related to the AoAs. In this framework, we suitably modify data distribution to account for these unobservable variables and estimate the unknown parameters by maximizing the modified distribution through the EM algorithm. This result represents the main technical innovation of this manuscript. Moreover, the estimates provided by the EM-based procedure are used to implement maximum a posteriori rules to classify the range bins with the highest posterior probability of containing target components with a given AoA. Doing so, we estimate targets' positions, AoAs, and number without exploring all the possible configurations of targets arising from the considered operation scenario. It is clear that the proposed strategy can significantly reduce the computational burden associated with maximum likelihood principle. Finally, we build up a Likelihood Ratio Test (LRT) that leverages the estimates provided by the EM-based procedure. The performance assessment, carried out over synthetic data, points out that the newly proposed detector demonstrates superior detection performance with respect to the considered conventional competitors, at least for the considered simulation parameters. Moreover, as a by product, it provides an indication of the targets' positions, AoA, and number.
 
 The remainder of this paper is organized as follows. Section II deals with the problem formulation and fundamental definitions required for the ensuing developments while Section III is devoted to detector designs. The estimation and detection performances of the new detector are assessed in Section IV. Finally, in Section V, the conclusions and some hint for future research are provided.

\subsection{Notation}

In the sequel, vectors and matrices are denoted by boldface
lower-case and upper-case letters, respectively. The symbols $\det(\cdot)$, $\tr(\cdot)$, $(\cdot)^T$, $(\cdot)^{*}$, $(\cdot)^\dag$ and $(\cdot)^{-1}$ denote the determinant, trace, transpose, conjugation, conjugate transpose, and inverse, respectively, and symbol $\boldsymbol{\nu}^T(\cdot)$ represents the vector-valued function selecting the distinct entries of a matrix. $\C^{M\times N}$ is the Euclidean space of $(M\times N)$-dimensional complex-valued matrices (or vectors if $N=1$). The 
modulus of a real number $a$ is denoted by $|a|$. $\bI$ and $\bzero$ stand for the identity matrix and the null vector or matrix of proper size, $A \backslash B$ represents the difference set between sets $A$ and $B$, while $A \cup B$ denotes the union of sets $A$ and $B$. Given a vector $\textbf{a}$,
$\diag(\textbf{a})$ indicates the diagonal matrix whose $i$th diagonal
element is the $i$th entry of $\textbf{a}$. The symbol $\Rightarrow$ means that the left-hand side implies the right-hand side. Finally, we write $\bx\sim\cC\cN_{N}(\bm, \bM)$ to indicate that $\bx$ is an $N$-dimensional complex  normal random vector  with mean $\bm$ and positive definite covariance matrix $\bM$. 

\section{Problem Formulation and preliminary definitions}
\label{Sec:SignalModel}

Let us consider the homogeneous scenario and the problem of detecting multiple  point-like targets in the presence of Gaussian disturbance. The returns from $K$ range bins are collected by a uniform linear array of $N>1$ sensors to form $N$-dimensional vectors $\bz_k\in\C^{N\times1}$, $k\in\Omega =\{1,\ldots, K\}$, with $\Omega$ standing for the index set of the $\bz_k$s. Assume that multiple point-like targets are present in the region of interest and that the indices of the range bins contaminated by target components, i.e., $\Omega_{T}=\{k_1,\ldots,k_T\}$ with $T\geq1$ the number of targets, are unknown. Then, the set of data vectors free of signal components is defined as $\Omega_{I}=\Omega \backslash \Omega_{T}$. The detection problem under consideration can be formulated as the following binary hypothesis test 

\be
\left\{
\begin{array}{l}
	\begin{aligned}
		&H_{0}: 
		\bz_{k} = \bn_{k} \sim\cC\cN_{N}(\bzero,\bM), \ k  \in \Omega,\\
		&H_{1}: \left\{
		\begin{array}{l}
			\bz_{k} = \alpha_{k} \bv(\theta_k) + \bn_{k} \sim\cC\cN_{N}(\alpha_{k} \bv(\theta_k),\bM), \\ \ \ \ \ \ \ \ \ \  \ \ \ \ \ \ \ \ \ \ \ \ \ \ \ \ \ \ \ \ \ \ \ \ \ \ \ \ \ \ \ \ \ k \in \Omega_{T},\\
			\bz_{k} = \bn_{k} \sim\cC\cN_{N}(\bzero,\bM), \ k \in \Omega_{I},
		\end{array}
		\right.
	\end{aligned}
\end{array}
\right.
\label{Eq1}
\ee
where $H_0$ and $H_1$ denote the interference-only and the signal-plus interference hypotheses, $\bn_{k} \in \C^{N\times1}, k\in \Omega$, are Independent and Identically Distributed (IID) interference components comprising thermal noise and clutter,\footnote{All the range bins are supposed to share the same statistical characterization of the interference.} $\bM \in \C^{N\times N}$ is the positive definite interference covariance matrix, $\alpha_{k} \in \C$ and $\bv(\theta_k)\in \C^{N\times 1}$ are the deterministic target amplitude and spatial steering vector in the $k$th range cell, respectively. Finally, $\theta_k$, $k \in \Omega_{T}$, are the unknown AoAs of the targets.

In realistic scenarios, problem defined in \eqref{Eq1} is difficult due to the information lack about targets' steering vectors, targets' number, as well as the positions of multiple targets within the $K$ range bins. Solving these issues by means of the conventional maximum likelihood approach is a formidable task. For this reason, we introduce an approach based upon a modification of the LVM. Specifically, under $H_1$, we first assume the presence of  hidden random variables that classify the $\bz_k$s according to the presence of a possible target.
To this end, we denote by  $c_k, k\in \Omega$, IID random variables with unknown Probability Mass Function (PMF) $P(c_k=s) = \pi_s$ that satisfies $\sum_{s=0}^1 \pi_s =1$. 
These random variables are such that when $c_k=1$ (namely, a target is present in the $k$th range bin), 
then $\bz_k\sim \cC\cN_N(\alpha_k\bv(\theta_k),\bM)$, whereas if $c_k=0$ (i.e., the $k$th range bin does not contain
target components), then $\bz_k\sim \cC\cN_N(\bzero,\bM)$. The inner LVM is related to the AoAs and consists in
introducing other IID (hidden) discrete random variables $e_k$, $k\in\Omega$,
whose unknown PMF is defined as $P(e_k=\theta_n)=p_n$ with
$\theta_n \in \Omega_{\theta}=\{\theta_1,\ldots, \theta_{K_{\theta}}\}$ and $\sum_{n=1}^{K_{\theta}} p_n=1$, to describe the
probability that a target (case $c_k=1$) originates from the angle $\theta_n$ in the $k$th range bin.
Moreover, $\Omega_{\theta}$ represents a suitable sampling grid covering the 
angular sector of interest (such as that spanned by the $3$ dB width of the mainbeam)
and $n \in \cA=\{1,\ldots,K_{\theta}\}$ with $K_{\theta}$ the number of grid angles.
Under this assumption, the Probability Density Function (PDF) of $\bz_k$ when $c_k=1$ can be further 
recast as\footnote{Recall that only one element in $\Omega_{\theta}$ represents the 
	real target AoA within a range cell. The AoA does not make sense when the target 
	is not present in that range cell, namely $c_k=0$.}
\be
f(\bz_k|c_k=1;\Psi_{k,1})=\sum_{n=1}^{K_{\theta}}f(\bz_k|c_k=1,e_k=n;\Phi_{k,n})p_n,
\ee
where
\begin{multline}
	f(\bz_k|c_k=1,e_k=n;\Phi_{k,n})
	\\
	=\frac{\exp\{-\tr[\bM^{-1}(\bz_k-\alpha_{k,n}\bv(\theta_n))(\bz_k-\alpha_{k,n}\bv(\theta_n))^\dag]\}}
	{\pi^N\det(\bM)}
	\label{PDF_h1}
\end{multline}
and $\Psi_{k,1}=\cup_{n=1}^{K_{\theta}}\Phi_{k,n}$ with $\alpha_{k,n}\in\C$ the 
response of the target component originating from $\theta_n$ in the $k$th range bin \footnote{Note that $\alpha_{k,n}$ is the target amplitude associated with $e_k=n$ and replace the $\alpha_k$ of \eqref{Eq1} to make it possible that the value of the target amplitude changes with $n$.},
$\Phi_{k,n}=\{ \alpha_{k,n},\theta_n,\bnu(\bM) \}$, and $\bnu(\bM)$ is a vector
whose entries are the generally distinct elements of $\bM$.

Gathering the above results, the hierarchical LVM leads to the following PDF for $\bz_k$ under $H_1$
\begin{align}
	f(\bz_k;\Psi_{k})&=\pi_1f(\bz_k|c_k=1;\Psi_{k,1})
	+\pi_0 f(\bz_k|c_k=0;\Psi_{k,0})
	\nonumber
	\\
	&=\pi_1\sum_{n=1}^{K_{\theta}}f(\bz_k|c_k=1,e_k=n;\Phi_{k,n})p_n
	\nonumber
	\\
	&+\pi_0 f(\bz_k|c_k=0;\Psi_{k,0}),
\end{align}
where $\Psi_{k,0}=\{\bnu(\bM)\}$($\subset \Psi_{k,1}$), $\Psi_{k}=\Psi_{k,0}\cup \Psi_{k,1}$, and
\be
f(\bz_k|c_k=0;\Psi_{k,0})=\frac{\exp\{-\tr[\bM^{-1}\bz_k\bz_k^\dag]\}}
{\pi^N\det(\bM)}.
\label{PDF_h0}
\ee

\section{estimation procedure and detection architecture}
In the context of multiple point-like targets detection distributed in range and
azimuth, problem \eqref{Eq1} is challenging since the $\alpha_{k,n}$s, $\theta_k$s, $\bM$, $T$, and $\Omega_{T}$ are unknown and must be estimated by the exploiting the available observations. Actually, the main critical issues arising from this problem are the following: one aspect involves the classification of the target echoes across the range bins as well as the estimation of the corresponding AoAs within the angular sector of interest; the other aspect involves the design of a detection architecture based on the aforementioned classification results. Therefore, we devise an unsupervised classifier by using the ``hierarchical" LVM described in Section II and the EM algorithm. Then, we design a novel adaptive decision scheme relying on an ad hoc LRT.

\subsection{Unsupervised classifier based on hierarchical LVM and EM algorithm}
\label{subsectionA}

In what follows, we describe an optimization procedure based on the EM algorithm to come up with suitable estimates of the unknown parameters.\footnote{It is worth noticing that closed-form expressions for the estimates of $\alpha_{k,n}, \bM, \pi_s$, and $p_n$, $s=0,1$, $n \in \cA$, $k\in \Omega$, can be obtained using the proposed EM-based estimation procedure, while the estimates of $T$, $\Omega_{T}$, and $\theta_n$ are obtained by applying a suitable maximum \textit{a posteriori} rule (a point better explained below). } Before proceeding with the design, we denote by $\bZ=[\bz_1, \ldots, \bz_K] \in\C^{N \times K}$ the overall data matrix. Then, exploiting the Jensen’s Inequality \cite{murphy2012machine, 10645068}, the joint log-likelihood of $\bZ$ under the $H_1$ hypothesis satisfies the following inequalities, 
\begin{align}
	L(\bZ;\Psi) &= \sum_{k=1}^{K} \log{ \sum_{s=0}^1 f(\bz_k|c_k=s;\Psi_{k,s}) \pi_s},\nonumber \\ 
	& \geq \sum_{k=1}^{K} \sum_{s=0}^1 q_k(s) \log{\frac{f(\bz_k|c_k=s;\Psi_{k,s}) \pi_s}{q_k(s)}}\nonumber \\
	& \geq \sum_{k=1}^{K} \left \{ q_k(0) \log{\frac{f(\bz_k|c_k=0;\Psi_{k,0}) \pi_0}{q_k(0)}} \right.\nonumber\\
	&\left.+ q_k(1) \sum_{n=1}^{K_{\theta}} r_k(n)\log{ \frac{f(\bz_k|c_k=1,e_k=n;\Phi_{k,n})p_n}{r_k(n)}} \right. \nonumber\\
	&\left. + q_k(1) \log{\frac{\pi_1}{q_k(1)}} \right \},
	\label{Eq5}
\end{align}
where the first inequality becomes equality when $q_k(s)$ is proportional to $f(\bz_k|c_k=s;\Psi_{k,s})\pi_s$ and the second inequality becomes equality when $r_k(n)$ is proportional to $f(\bz_k|c_k=1,e_k=n;\Phi_{k,n})p_n$. In \eqref{Eq5}, we denote by $\Psi=\cup_{k=1}^{K} \Psi_k$, while by $q_k(s)$ and $r_k(n)$ the \textit{a posteriori} probabilities of the events $c_k=s$ given $\bz_k$ and $e_k=n$ given $\bz_k$ and $c_k=1$, respectively. The E-step in the EM algorithm leads to the following updates 
	\begin{align}
	q_k^{(m-1)}(0)  = \frac{F_0^{(m-1)} \widehat{\pi}_0^{(m-1)}  } { F_0^{(m-1)} \widehat{\pi}_0^{(m-1)} + F_1^{(m-1)} \widehat{\pi}_{1}^{(m-1)} }, 
	\label{q0_0} 
   \end{align}
\be
q_k^{(m-1)}(1) =  \frac{ F_1^{(m-1)} \widehat{\pi}_1^{(m-1)}  } { F_0^{(m-1)}\widehat{\pi}_0^{(m-1)} + F_1^{(m-1)} \widehat{\pi}_{1}^{(m-1)} }, 
\label{q1_0}  
\ee
\be
r_k^{(m-1)}(n) = \frac{f\left(\bz_k|c_k=1,e_k=n;\widehat{\Phi}_{k,n}^{(m-1)}\right)  \widehat{p}_n^{(m-1)} } { F_1^{(m-1)} },
\label{rn_0} 
\ee
where $F_0^{(m-1)} = f\left(\bz_k|c_k=0;\widehat{\Psi}_{k,0}^{(m-1)} \right)$, $F_1^{(m-1)} = \sum_{n=1}^{K_{\theta}} f\left(\bz_k|c_k=1,e_n=n;\widehat{\Phi}_{k,n}^{(m-1)}\right) \widehat{ p}_{n}^{(m-1)}$, with $\widehat{\Psi}_{k,s}^{(m-1)}$, $\widehat{\pi}_s^{(m-1)}$, and $\widehat{p}_{n}^{(m-1)}$, $s=0,1$, $n \in \cA$, the estimates of the unknown parameters  at the $(m-1)$th iteration. In order to control the overestimation of target components, we introduce a penalization in \eqref{q0_0}-\eqref{q1_0} represented by the function $e^{u(s,\rho)} = e^{\frac{(N^2+3s)(1+\rho)}{2}}, s=0,1, \rho \geq 1$, inspired by the Generalized Information Criterion (GIC) \cite{9444189, 10058041}. As a consequence, the modified expressions are given by

\be
q_k^{(m-1)}(0) = \frac{F_0^{(m-1)} \widehat{\pi}_0^{(m-1)}  e^{-u(0,\rho)}} { F_0^{(m-1)} \widehat{\pi}_0^{(m-1)}e^{-u(0,\rho)} + F_1^{(m-1)} \widehat{\pi}_{1}^{(m-1)} e^{-u(1,\rho)} },  
\ee

\be
q_k^{(m-1)}(1) =  \frac{F_1^{(m-1)} \widehat{\pi}_1^{(m-1)}  e^{-u(1,\rho)}} {F_0^{(m-1)} \widehat{\pi}_0^{(m-1)} e^{-u(0,\rho)} + F_1^{(m-1)} \widehat{\pi}_{1}^{(m-1)} e^{-u(1,\rho)}}, 
\ee


The M-step allows us to obtain the updates of the unknown parameter estimates.  Ignoring the irrelevant terms, we can write the maximization problem as 
\begin{align}
  &\max_{ {\bf \balpha}_{k}, {\bf M} , \pi_s,s=0,1,\atop p_n, n \in\cA, k\in \Omega }  \sum_{k=1}^{K} \left\{q_k^{(m-1)}(0) \log{f(\bz_k|c_k=0;\Psi_{k,0})} \right. \nonumber \\
	& \left.+ q_k^{(m-1)}(1) \sum_{n=1}^{K_{\theta}} r_k^{(m-1)}(n)\log{ f(\bz_k|c_k=1,e_k=n;\Phi_{k,n})} \right. \nonumber \\
	&\left. +\sum_{s=0}^1 q_k^{(m-1)}(s) \log{\pi_s} + q_k^{(m-1)}(1) \sum_{n=1}^{K_{\theta}} r_k^{(m-1)}(n) \log{p_n}\right\},
	\label{M-step}
\end{align}
 where $\balpha_{k} = [\alpha_{k,1},\ldots,\alpha_{k,K_{\theta}}]^{T} \in \C^{K_{\theta}\times1}$. Observe that the optimizations with respect to $\pi_s,s=0,1$, and $ p_n, n \in \cA$, are independent of that over $\balpha_{k}$ and $\bM$, and can be solved using the Lagrange multipliers method \cite{2021Learning, 10645068}. Specifically, starting from the $\pi_s$s, it is not difficult to show that the solution of the problem
\be
\left\{
\begin{array}{l}
	\begin{aligned}
		& \max_{\pi_s, s=0,1} \left\{\sum_{k=1}^{K}\sum_{s=0}^1 q_k^{(m-1)}(s) \log{\pi_s} \right\},\\
		& \sum_{s=0}^1 \pi_s =1,
	\end{aligned}
\end{array}
\right.
\ee
is
\be
\widehat{\pi}_s^{(m)}=\sum_{k=1}^{K}q_k^{(m-1)}(s) /K, \ s=0,1.
\ee
As for the $p_n$s, the related problem is
\be
\left\{
\begin{array}{l}
	\begin{aligned}
		& \max_{p_n, n \in \cA} \sum_{k=1}^{K} \left\{
		q_k^{(m-1)}(1) \sum_{n=1}^{K_{\theta}} r_k^{(m-1)}(n) \log{p_n} \right\},\\
		& \sum_{n=1}^{K_{\theta}} p_n =1.
	\end{aligned}
\end{array}
\right.
\label{Laglange2}
\ee
And the corresponding Langrangian function  is
\be
\sum_{k=1}^{K}
q_k^{(m-1)}(1) \sum_{n=1}^{K_{\theta}} r_k^{(m-1)}(n) \log{p_n} -\lambda\left(\sum_{n=1}^{K_{\theta}} p_n -1\right),
\ee
where $\lambda>0$ is the Lagrange multiplier. Setting to zero its first derivative with respect to $p_n$ leads to
\be
p_n = \frac{1}{\lambda} \sum_{k=1}^{K}
q_k^{(m-1)}(1) r_k^{(m-1)}(n).
\ee
Exploiting the constraint, we obtain
\begin{align}
\lambda &= \sum_{k=1}^{K}
q_k^{(m-1)}(1) \sum_{n=1}^{K_{\theta}} r_k^{(m-1)}(n) = \sum_{k=1}^{K}
q_k^{(m-1)}(1)\ \nonumber \\ 
&\Rightarrow  \widehat{p}_n^{(m)}=\frac{\sum_{k=1}^{K}q_k^{(m-1)}(1)r_k^{(m-1)}(n) }{\sum_{k=1}^{K}q_k^{(m-1)}(1)}, \ n \in \cA.
\end{align}
The final maximization is with respect to $\balpha_{k}$ and $\bM$. To this end, plugging \eqref{PDF_h0} and \eqref{PDF_h1} into \eqref{M-step}, the related
objective function can be written as
\begin{align}
& \max_{{\bf \balpha}_{k},{\bf M}, k\in \Omega} \sum_{k=1}^{K} \left\{q_k^{(m-1)}(0) \log{f(\bz_k|c_k=0;\Psi_{k,0})} \right. \nonumber\\
&\left.+ q_k^{(m-1)}(1) \sum_{n=1}^{K_{\theta}} r_k^{(m-1)}(n)\log{ f(\bz_k|c_k=1,e_k=n;\Phi_{k,n})} \right\}, \nonumber\\
	& = -KN\log{\pi}-K\log{\det(\bM)} \nonumber\\
	& -\tr\left\{ \bM^{-1} \sum_{k=1}^{K} q_k^{(m-1)}(0) \bS_k \right\}  \nonumber\\
	& -\tr\left\{\bM^{-1}\sum_{k=1}^{K} \sum_{n=1}^{K_{\theta}} q_k^{(m-1)}(1) r_k^{(m-1)}(n) \bS_{n,k}  \right\} ,\nonumber\\
	& = -KN\log{\pi}-K\Bigg\{\log{\det(\bM)}    \nonumber\\
	& +\tr\left[\bM^{-1} 
	\sum_{k=1}^{K} \Big(q_k^{(m-1)}(0)\bS_k  \right.  \nonumber\\
&  \left. 	+\sum_{n=1}^{K_{\theta}}q_k^{(m-1)}(1) r_k^{(m-1)}(n)\bS_{n,k}\Big)/K\right] \Bigg\},\nonumber\\
	& \leq -KN\log{\pi} - K\log\det\left\{ \sum_{k=1}^{K} \Big(q_k^{(m-1)}(0)\bS_k \right. \nonumber\\
& \left.  +\sum_{n=1}^{K_{\theta}}q_k^{(m-1)}(1) r_k^{(m-1)}(n)\bS_{n,k}\Big)/K\right\}-KN,
	\label{MaximizationM}
\end{align}
where $\bS_k=\bz_k\bz_k^{\dagger}\in \C^{N\times N}$, $\bS_{n,k}=(\bz_k-\alpha_{k,n}\bv(\theta_n))(\bz_k-\alpha_{k,n}\bv(\theta_n))^{\dag}\in \C^{N\times N}$, and the inequality is due to $\det(\bA)e^{-\tr\left(\bA\right)}\leq e^{-N}$ for  $\bA\in\C^{N \times N}$  any matrix with nonnegative eigenvalues \cite{mirsky1959trace, lutkepohl1997handbook}. Hence, the estimate update for $\bM$  is given by
\be
\widehat{\bM}^{(m)} = \frac{\displaystyle \sum_{k=1}^{K} \left(q_k^{(m-1)}(0)\bS_k+q_k^{(m-1)}(1)\displaystyle \sum_{n=1}^{K_{\theta}} r_k^{(m-1)}(n)\bS_{n,k}\right)}{K}. 
\ee


Replacing $\widehat{\bM}^{(m)}$ in \eqref{MaximizationM} and neglecting the irrelevant constants, the maximization with respect to $\balpha_{k}$ is equivalent to
\begin{align}
&	\min_{\alpha_{k,n}\atop n \in \cA,k \in \Omega}  K\log\det\left\{ \sum_{k=1}^{K} \Big(q_k^{(m-1)}(0)\bS_k\right. \nonumber \\
	&\left. + q_k^{(m-1)}(1)\sum_{n=1}^{K_{\theta}} r_k^{(m-1)}(n)\bS_{n,k}\Big)/K\right\}\nonumber \\
	&\Rightarrow \min_{\alpha_{k,n}\atop n \in \cA,k \in \Omega} \det\left\{  \sum_{k=1}^{K} \Big( q_k^{(m-1)}(0)\bS_k\right. \nonumber \\
	&+q_{k}^{(m-1)}(1) r_{k}^{(m-1)}(n)\bS_{n,k}  \nonumber \\
	& \left.+ q_{k}^{(m-1)}(1) \sum_{n^{\prime}=1, \atop n^{\prime}\neq n}^{K_{\theta}} r_{k}^{(m-1)}(n^{\prime})\bS_{n^{\prime},k} \Big) / K \right\},
	\label{EQ18}
\end{align}
with $\bS_{n^{\prime},k}=\left(\bz_{k}-\alpha_{k,n^{\prime}}\bv(\theta_{n^{\prime}})\right)\left(\bz_{k}-\alpha_{k,n^{\prime}}\bv(\theta_{n^{\prime}})\right)^{\dag}$. It should be noted that the joint maximization with respect to the $\alpha_{k,n}$s is difficult from a mathematical point of view. Therefore, we resort to an iterative maximization with respect to $\overline{\balpha}_n=[\alpha_{1,n},\ldots,\alpha_{K,n}]^{T} \in \C^{K\times 1}$. Specifically, we assume that $\overline{\balpha}_i, i \in \cA, i \neq n$, are known and maximize the objective function respect to $\overline{\balpha}_n$. This procedure is repeated $\forall n \in \cA$. To this end, let us define
\begin{align}
&\bB_n^{(m-1)} \nonumber \\ 
&= \sum_{k=1}^{K} q_k^{(m-1)}(0)\bS_k+ \sum_{k=1}^{K}q_{k}^{(m-1)}(1)\sum_{n^{\prime}=1,  \atop n^{\prime}\neq n}^{K_{\theta}} r_{k}^{(m-1)}(n^{\prime})\bS_{n^{\prime},k},
\end{align}
and 
\begin{align}
&\bX_n = \left[\sqrt{Q_{1,n}^{(m-1)}}\bz_1, \ldots,\sqrt{Q_{K,n}^{(m-1)}}\bz_K\right], \nonumber \\ & \bbeta_n = \left[\sqrt{Q_{1,n}^{(m-1)}}\alpha_{1,n}, \ldots,\sqrt{Q_{K,n}^{(m-1)}}\alpha_{K,n}\right]^{T},
\end{align}
with $Q_{k,n}^{(m-1)}=q_{k}^{(m-1)}(1) r_{k}^{(m-1)}(n) > 0, k \in \Omega$. Then, the maximization with respect to $\overline{\balpha}_n$ assuming that $\overline{\balpha}_i, i \in \cA, i\neq n$, are known is tantamount to
\begin{align}
&\min_{\bbeta_n, n\in \cA}\left[\frac{1}{K}\right]^{N} \det\left[\bB_n^{(m-1)} + \bH_A\bH_A^{\dag}\right],\nonumber\\
=&\min_{\bbeta_n, n\in \cA} \left[\frac{1}{K}\right]^{N} \det[\bB_n^{(m-1)}] \det[\bI+\bH_A^{\dag} (\bB_n^{(m-1)})^{-1}\bH_A ] .
\end{align}
where $\bH_A =\bX_n-\bv(\theta_n)\bbeta_n^{T}$.
Neglecting the irrelevant items, the above optimization problem can be written as 
\begin{align}
 &\min_{\bbeta_n, n \in \cA}  \det\left[\bI+ \bX_n^{\dag}\left(\bB_n^{(m-1)}\right)^{-1}\bX_n \right. \nonumber\\
&\left. - \bX_n^{\dag}\left(\bB_n^{(m-1)}\right)^{-1}\bv(\theta_n)\bbeta_n^{T}
 -\bbeta_n^{*}\bv(\theta_n)^{\dag}\left(\bB_n^{(m-1)}\right)^{-1}\bX_n\right. \nonumber\\ &\left.+\bbeta_n^{*}\bv(\theta_n)^{\dag}\left(\bB_n^{(m-1)}\right)^{-1}\bv(\theta_n) \bbeta_n^{T} \right], \nonumber \\
& = \min_{\beta_n,  n \in \cA} \det\left[\bI+ \bX_n^{\dag}\left(\bB_n^{(m-1)}\right)^{-1}\bX_n \right. \nonumber\\
&\left.- \bX_n^{\dag}\left(\bB_n^{(m-1)}\right)^{-1}\bv(\theta_n)A^{-1}A\bbeta_n^{T}\right. \nonumber\\
&\left.-\bbeta_n^{*}AA^{-1}\bv(\theta_n)^{\dag}\left(\bB_n^{(m-1)}\right)^{-1}\bX_n +\bbeta_n^{*} A A^{-1}A \bbeta_n^{T}\right. \nonumber\\
&\left.
 +\bX_n^{\dag}(\bB_n^{(m-1)})^{-1}\bv(\theta_n)\bv(\theta_n)^{\dag}\left(\bB_n^{(m-1)}\right)^{-1}\bX_n /A \right. \nonumber\\
 &\left.-\bX_n^{\dag}\left(\bB_n^{(m-1)}\right)^{-1}\bv(\theta_n)\bv(\theta_n)^{\dag}\left(\bB_n^{(m-1)}\right)^{-1}\bX_n / A \right], \nonumber\\
& = \min_{\beta_n, n\in \cA} \det\left[\bI+ \bX_n^{\dag}\left(\bB_n^{(m-1)}\right)^{-1}\bX_n + \bH_B A^{-1}\bH_B^{\dag}\right. \nonumber\\
&\left.-\bX_n^{\dag}(\bB_n^{(m-1)})^{-1}\bv(\theta_n)\bv(\theta_n)^{\dag}(\bB_n^{(m-1)})^{-1}\bX_n / A\right],
\end{align}
where $\bH_B =\bbeta_n^{*}A-\bX_n^{\dag}\left(\bB_n^{(m-1)}\right)^{-1}\bv(\theta_n)$, $A = \bv(\theta_n)^{\dag}(\bB_n^{(m-1)})^{-1}\bv(\theta_n)>0$. It is clear that the minimum is attained when
\be
(\widehat{\bbeta}_n^{T})^{(m)} = \frac{\bv(\theta_n)^{\dag}(\bB_n^{(m-1)})^{-1}\bX_n}{A}.
\ee
Thus, the update for $\overline{\balpha}_n$ can be finally written as
\be
(\widehat{\overline{\balpha}}_n^{T})^{(m)} = (\widehat{\bbeta}_n^{T})^{(m)} \bP_n^{(m-1)}, \ n \in \cA,
\label{estimate_alpha}
\ee
where $\bP_n^{(m-1)} = \diag \left\{(Q_{1,n}^{(m-1)})^{-1/2},\ldots,(Q_{K,n}^{(m-1)})^{-1/2}\right\}$. Notice that equation \eqref{estimate_alpha} is computed for each $n\in \cA$.

The whole cyclic procedure terminates when a stopping criterion is satisfied, such as when the relative variation of objective function 
is less than a suitable small positive number, namely, 
\be
\frac{\left|L(\bZ;\widehat{\Psi}^{(m)})-L(\bZ;\widehat{\Psi}^{(m-1)})\right|}{\left|L(\bZ;\widehat{\Psi}^{(m)})\right|} < \delta,
\label{Stop1}
\ee
or after a fixed number of iterations (a point better explained in Section \ref{sectionIV}), where $\widehat{\Psi}^{(m)}$ and $\widehat{\Psi}^{(m-1)}$ are the estimates of the parameter sets at the $(m)$th and $(m-1)$th iteration, respectively.

 Finally, the problems of classifying the $k$th range bin and estimating the AoA of a possible target can be accomplished by exploiting the following rules
\be
\widehat{s}_k = \arg\max_{s=0,1}  q_k^{(\overline{m}-1)}(s), \ k \in \Omega,
\label{ClassificationRule1}
\ee
and
\be
\widehat{n}_{k_1} = \arg\max_{n \in \cA}  r_{k_1}^{(\overline{m}-1)}(n), \ k_1\in \widehat{\Omega}_{T},
\label{ClassificationRule2}
\ee
where  $\overline{m}$ is the number of used iterations, and $\widehat{\Omega}_{T}=\{k \in \Omega: \widehat{s}_k=1  \}$ represents an estimate of the set of range bins containing target components. The final estimates of target AoAs are given by $\widehat{\theta}_{\widehat{n}_{k_1}}$, $k_1\in \widehat{\Omega}_{T}$, whereas the estimate of the target number, $\widehat{T}$ say, is the cardinality of $\widehat{\Omega}_{T}$.

\subsection{Adaptive detector based on LRT}

The estimates derived in Subsection \ref{subsectionA} can be used to build up an LRT-based detector. Specifically, replacing the parameters of interest 
with their estimates in the likelihood function, we obtain the following adaptive detector 
\begin{equation}
	\frac{f(\bZ;\widehat{\Psi}^{(\overline{m})},H_1) }{f(\bZ;\widehat{\bM}_0, H_0) } \test \eta,
	\label{blind_detector1}
\end{equation}
where $f(\bZ;\widehat{\Psi}^{(\overline{m})},H_1) = \prod_{k=1}^Kf(\bz_k;\widehat{\Psi}_{k}^{(\overline{m})})$ and $f(\bZ;\widehat{\bM}_0, H_0)= \prod_{k=1}^Kf(\bz_k;\widehat{\bM}_0)$ are the PDFs of $\bZ$ under $H_1$ and $H_0$, respectively. It is well-known that the maximum of the PDF of $\bz_k$
under $H_0$ with respect to $\bM_0$ is given by
\be
f(\bz_k;\widehat{\bM}_0) = \frac{\exp\left\{-\tr[\widehat{\bM}_0^{-1}\bz_k\bz_k^\dag]\right\}}
{\pi^N\det(\widehat{\bM}_0)},
\ee
with $\widehat{\bM}_0 = \bZ\bZ^{\dagger}/K$ the maximum likelihood estimate of $\bM$ under $H_0$ \cite{1336827, 8496824, 7376203 }. Then, decision scheme \eqref{blind_detector1} is statistically equivalent to
\begin{align}
&\sum_{k=1}^{K} \left\{\log\left(\widehat{\pi}_0^{(\overline{m})} f\left(\bz_{k}|c_{k}=0;\widehat{\Psi}_{k,0}^{(\overline{m})}\right) \right.\right. \nonumber \\
&\left.\left.+ \widehat{\pi}_1^{(\overline{m})} \sum_{n=1}^{K_{\theta}}f\left(\bz_{k}|c_{k}=1,e_{k} = n;\widehat{\Phi}_{k,n}^{(\overline{m})}\right) \widehat{p}_n^{(\overline{m})} \right) \right. \nonumber \\
&\left. - \log f(\bz_{k};\widehat{\bM}_0)\right\} \test \eta,
\end{align}
where $\widehat{\Psi}_{k,0}^{(\overline{m})}$ and $\widehat{\Phi}_{k,n}^{(\overline{m})}$ denote the final estimates of the sets $\Psi_{k,0}$ and $\Phi_{k,n}$, respectively.  



\section{Illustrative Examples and Discussion}
\label{sectionIV}

In this section, some numerical examples are presented to assess the behavior of the devised architecture in comparison with classical detectors that know the angular positions of the targets by using simulated data. Specifically, we assume that a
radar system equipped with $N=8$ spatial channels collects $K =24$ complex echoes from the region of interest. Within the angular sector $\Omega_{\theta}=[-20^{\circ}:2^{\circ}: 20^{\circ}]$, i.e., $K_{\theta} = 21$, three point-like targets appear in the 6th, 13th, and 16th range bins, respectively. The Signal-to-Interference-plus-Noise Ratio (SINR) is defined as SINR = $|\alpha_{k,i}|^2\bv(\theta_{k,i})^{\dag}\bM^{-1}\bv(\theta_{k,i})$,\footnote{Notice that the SINR does not depend on $k$ and $i$, thus each target shares the same SINR.} where $\alpha_{k,i}$ and $\theta_{k,i}$ are the response and angular location of the $i$th target in the $k$th range bin, $\bM = \sigma_n^2\bI+ \sigma_c^2 \bR_c$ with $\sigma_n^2=1$ and $\sigma_c^2 = 10^{\text{CNR}/10}\sigma_n^2$ the power of thermal noise  and clutter for Clutter-to-Noise Ratio (CNR) = 15 dB, respectively, while $\bR_c$ is the Hermitian positive definite matrix whose $(i,j)$th element is generated as $\rho_c^{|i-j|}, i, j = 1,\ldots,N$, and $\rho_c = 0.9$ \cite{10058041}. Accordingly, data are generated as follows
\be
\begin{array}{l}
	\left\{
		\begin{array}{l}
			\bz_{k}  \sim\cC\cN_{N}(\bzero,\bM), \ k  \in \Omega  \backslash \{6,13,16\},\\
			\bz_{k} \sim\cC\cN_{N}\left(\alpha_{6,3} \bv(\theta_{6,3}),\bM\right), \\ \bz_{k} \sim\cC\cN_{N}\left(\alpha_{13,13} \bv(\theta_{13,13}),\bM\right),\\ \bz_{k} \sim\cC\cN_{N}\left(\alpha_{16,17} \bv(\theta_{16,17}),\bM\right).
		\end{array}
	\right.
\end{array}
\ee

As for the initialization of the EM estimation procedure, we set  
\begin{align}
&\pi_{s}^{(0)} = 1/2, \ p_n^{(0)}= 1/K_{\theta}, \ \bM^{(0)} = \bZ\bZ^{\dag}/K, \nonumber \\
 &\alpha_{n,k}^{(0)}=\frac{\bv^{\dag}(\theta_n)\left(\bM^{(0)}\right)^{-1}\bz_k}{\bv^{\dag}(\theta_n)\left(\bM^{(0)}\right)^{-1}\bv(\theta_n)}, 
 \end{align}
 and $\rho = 3$, for $s=0,1$, $n\in \cA$, and $k \in \Omega$.

 \subsection{Matched Target AoAs}
\begin{figure}[tbp]
	\centering
	\includegraphics[width=.45\textwidth,height= 6.3 cm]{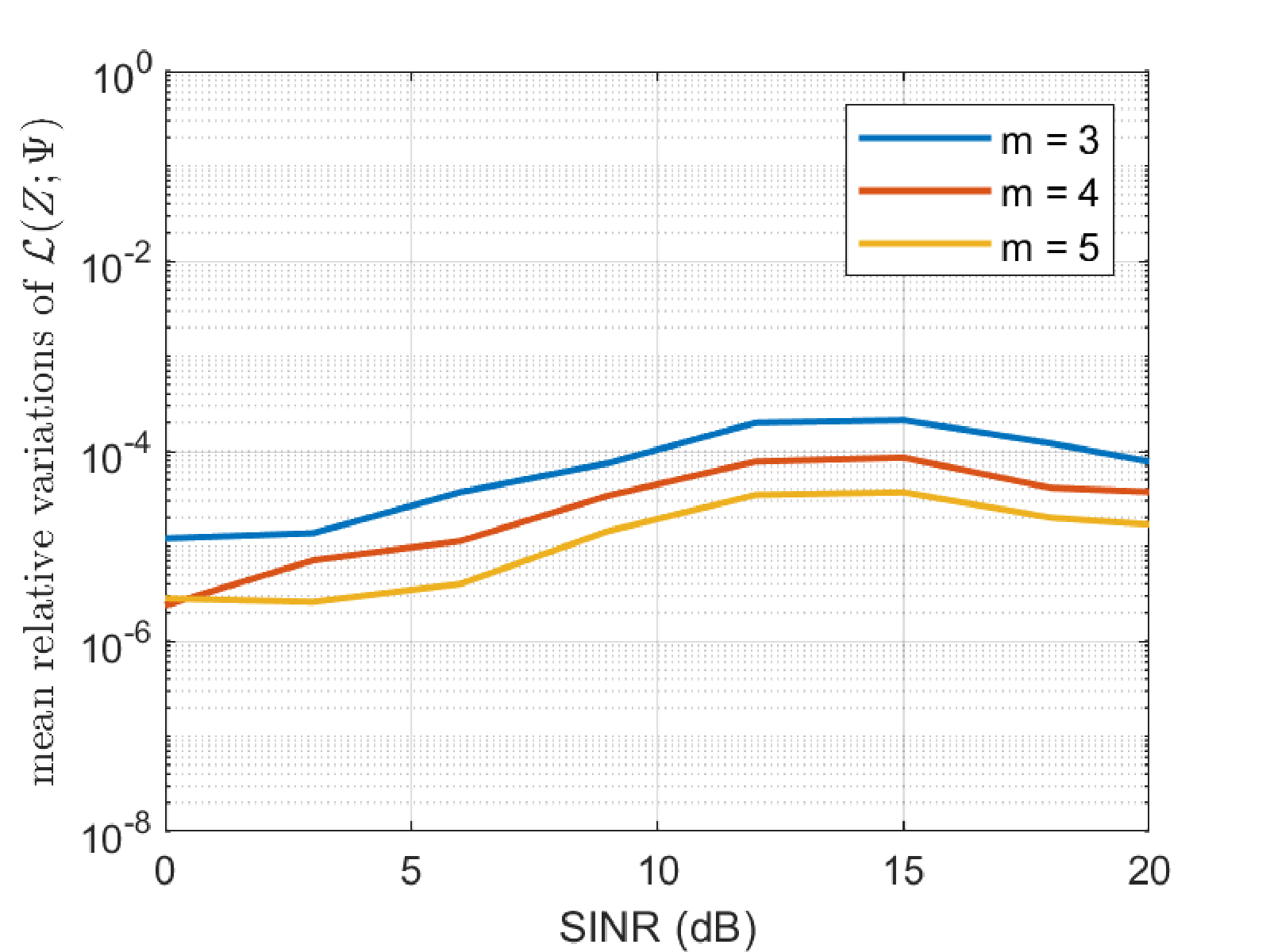}
	\caption{{ Mean relative variations of $L(\bZ;\Psi)$ versus SINR  for different values of $m$ and $t$ (matched AoAs).}}
	\label{convergence}
\end{figure}

 Firstly, we consider a scenario where the target AoAs are assumed to lie on the sampling grid of $\Omega_{\theta}$, as detailed in Table \ref{T1}. Before moving on,
 notice that a reliable convergence criterion is crucial to the EM iteration process, whose results play a decisive role in algorithm's performance. In Fig. \ref{convergence}, we plot the relative variations of the log-likelihood $L(\bZ;\Psi)$ as a function of the SINR
 for several values of $m$ averaged over 1000 Monte Carlo (MC) trials. This analysis is used to obtain a good balance between estimation quality and computational load. The results indicate that the cyclic optimization procedure based on EM converges to a variation lower than $10^{-4}$ when $m\geq4$. Thus, in the next illustrative examples, we select $\overline{m}=4$ to reduce the computational time.  

\begin{table}[!t]
	\begin{center}
		\caption{Parameter setting for multiple point-like targets with matched AoAs.}
		
		\begin{tabular}{cccccccc p{2 cm}}
			\toprule[1pt]
			\toprule[1pt]
			\textbf{Parameter Setting} & Target 1 & Target 2 & Target 3  \\
			\hline
			\renewcommand{\arraystretch}{2}
			\textbf{Rang bin} &6&  13   &   16 \\    
			\textbf{AoA($^\circ$)} &-16 &  4   &  12 \\         
			\toprule[1pt]
			\toprule[1pt]
		\end{tabular}
		\label{T1}
	\end{center}
\end{table} 

\begin{figure}[t!]
	\begin{center}
		\subfigure[ ]{\includegraphics[width=0.48\columnwidth]{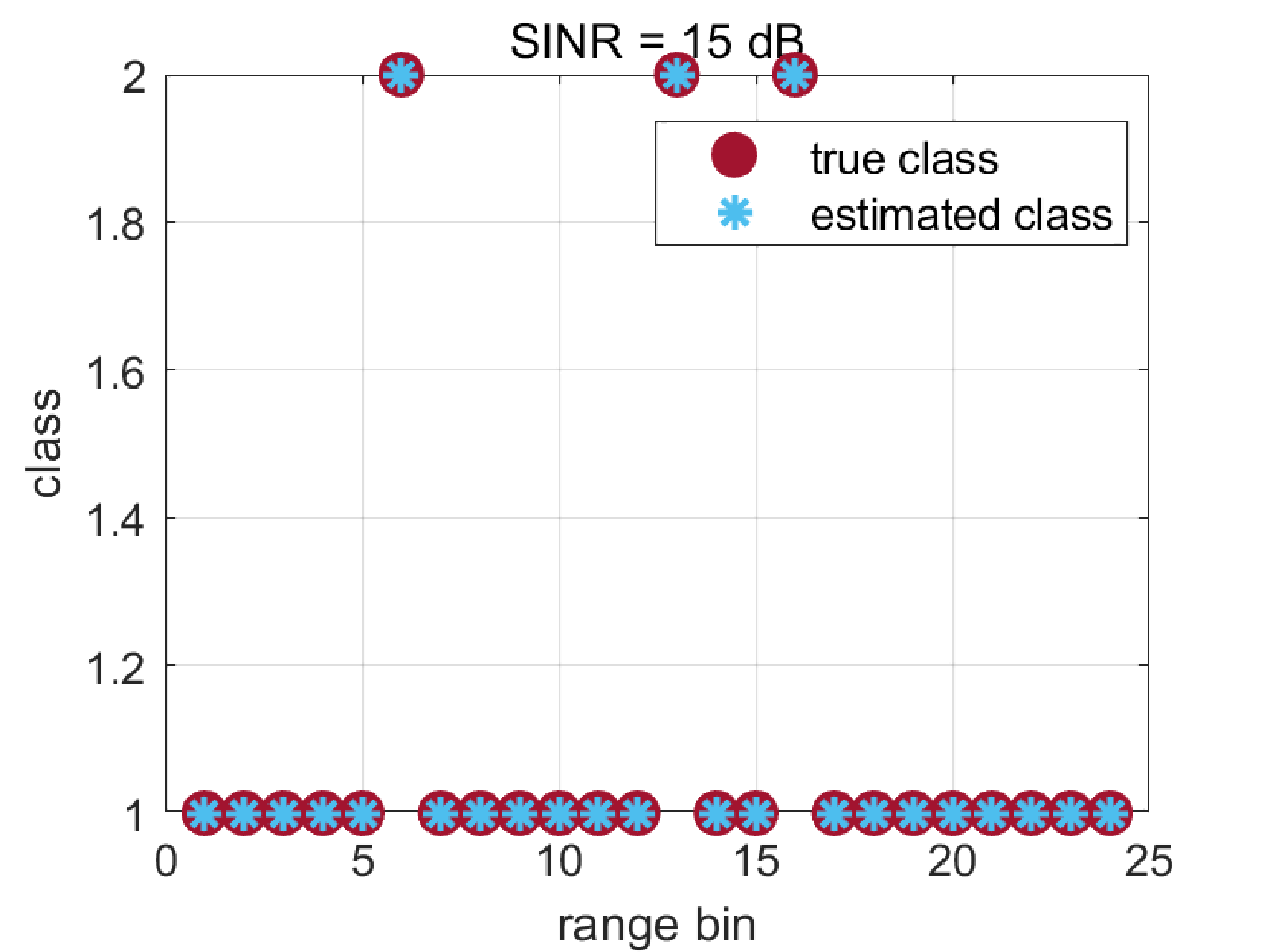}}
		\subfigure[]{\includegraphics[width=0.48\columnwidth]{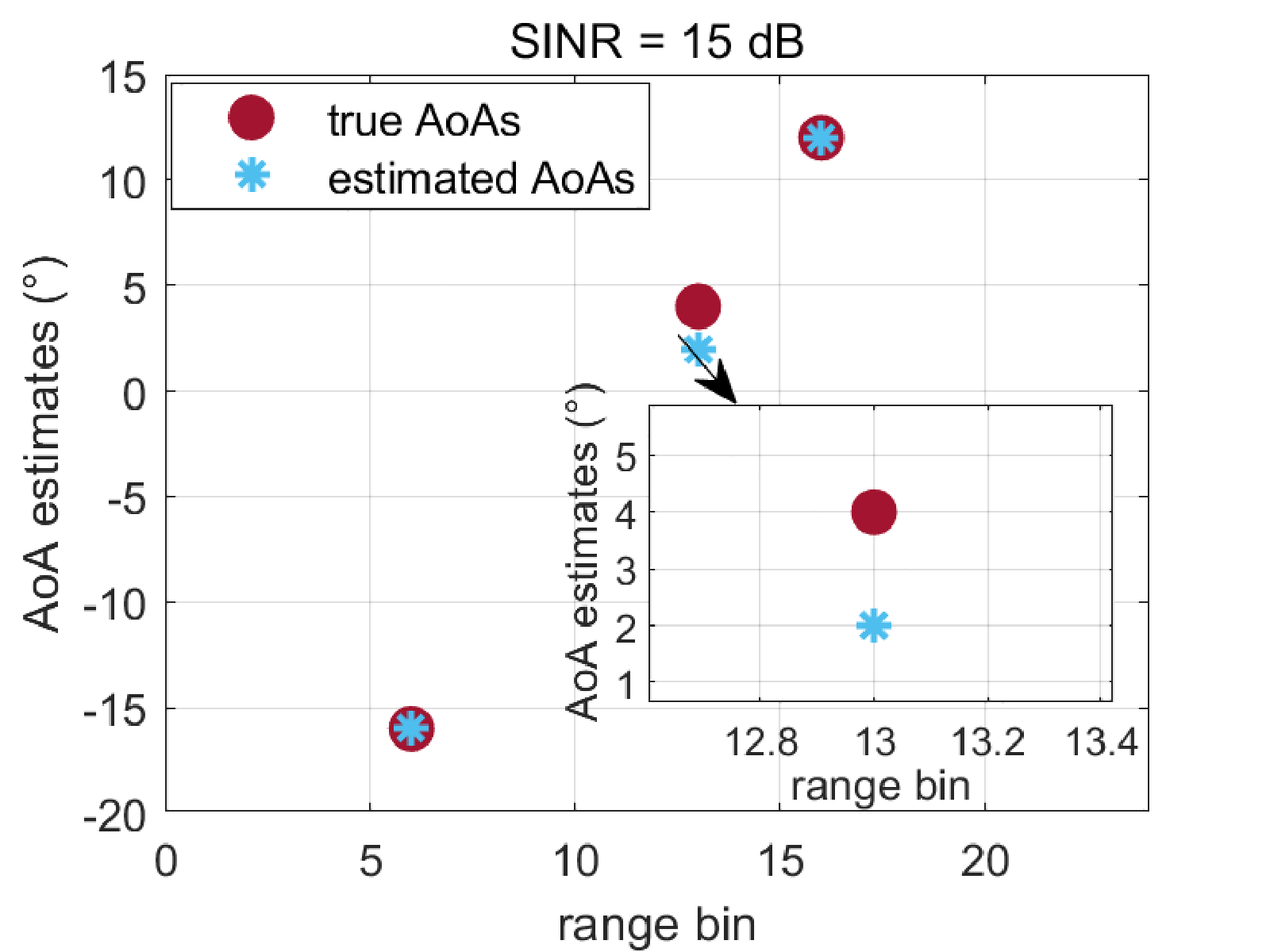}}\\
		\subfigure[]{\includegraphics[width=0.48\columnwidth]{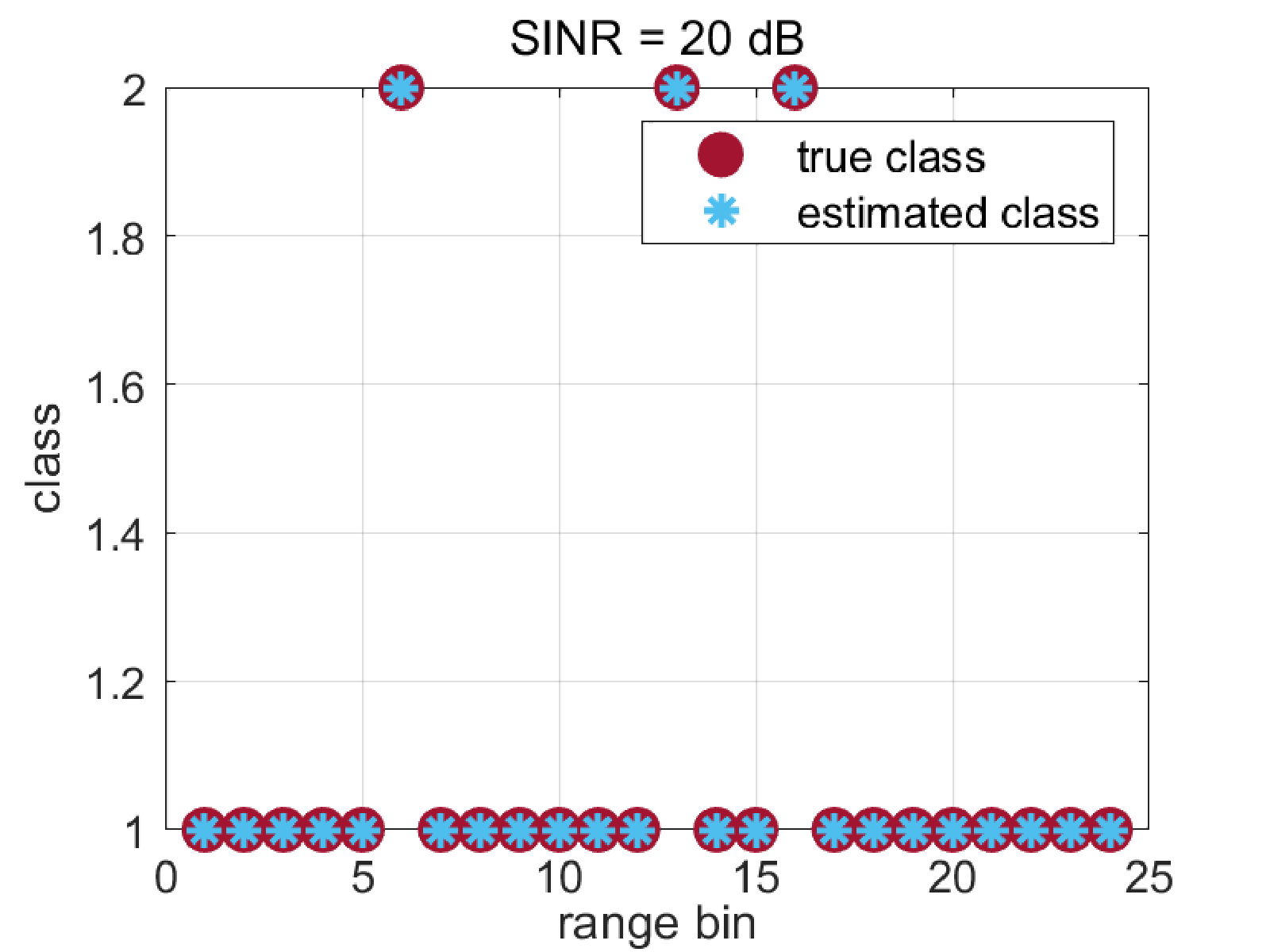}}
		\subfigure[]{\includegraphics[width=0.48\columnwidth]{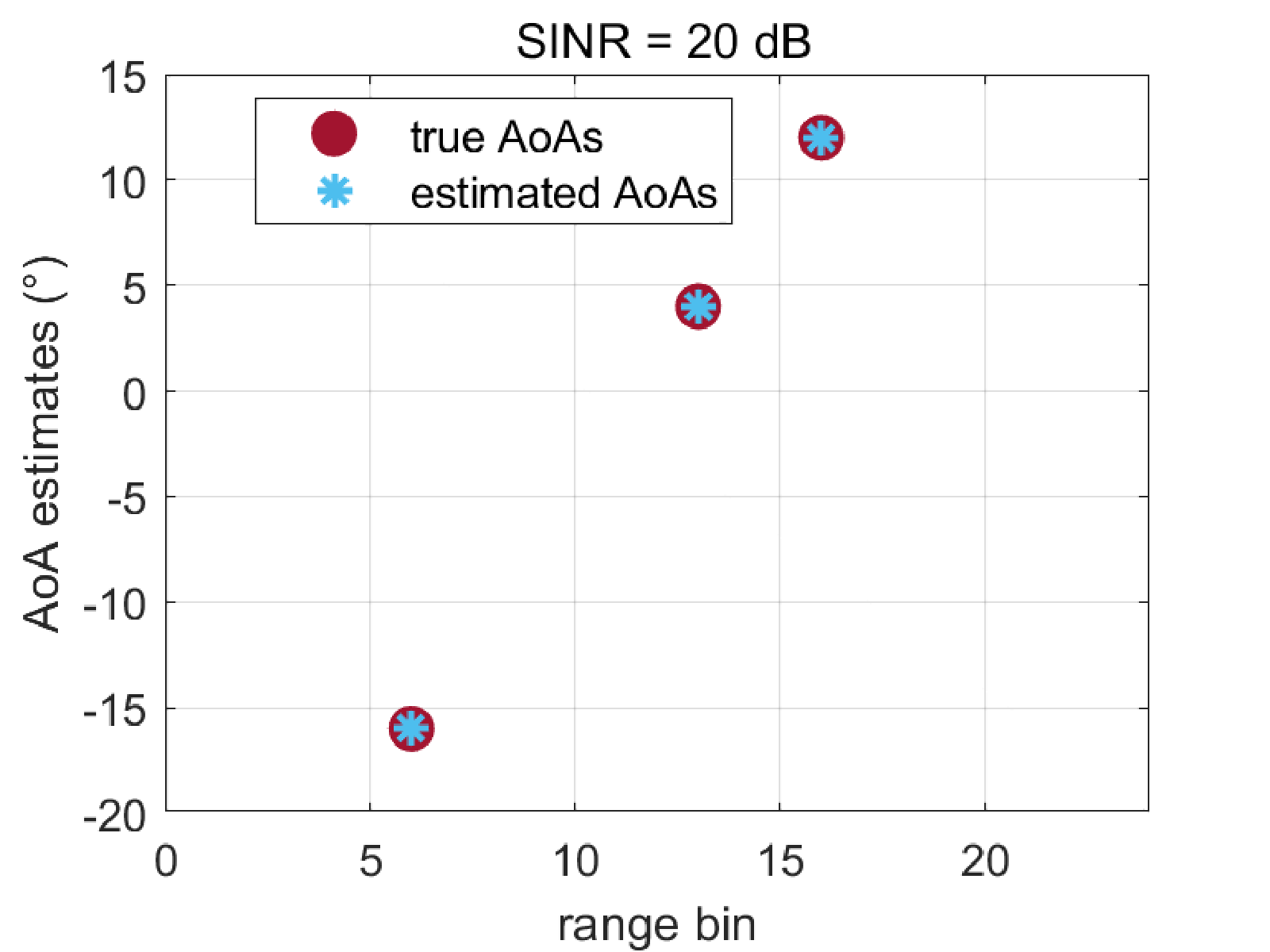}}\\
	\end{center}
	\caption{Classification snapshots over one trial (matched AoAs).
	}
	\label{snapshot}
\end{figure}

\begin{figure}[t!]
	\begin{center}
		{\includegraphics[width=.45\textwidth,height= 6.3 cm]{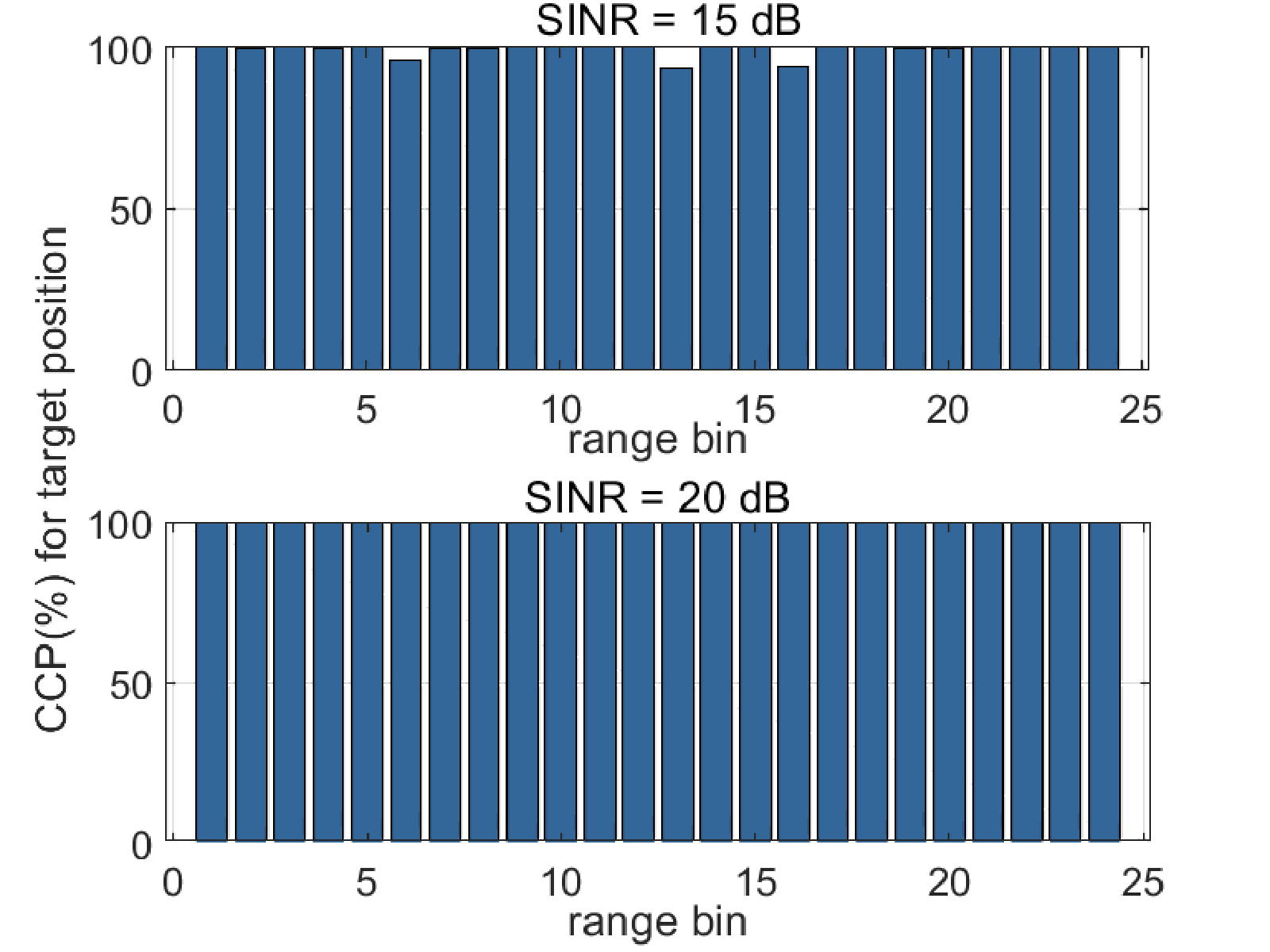}}\\
	\end{center}
	\caption{CCP ($\%$) for target position resorting to 1000 independent runs (matched AoAs). 
	}
	\label{CCP}
\end{figure}

\begin{figure}[t!]
	\begin{center}
		{\includegraphics[width=.45\textwidth,height= 6.3 cm]{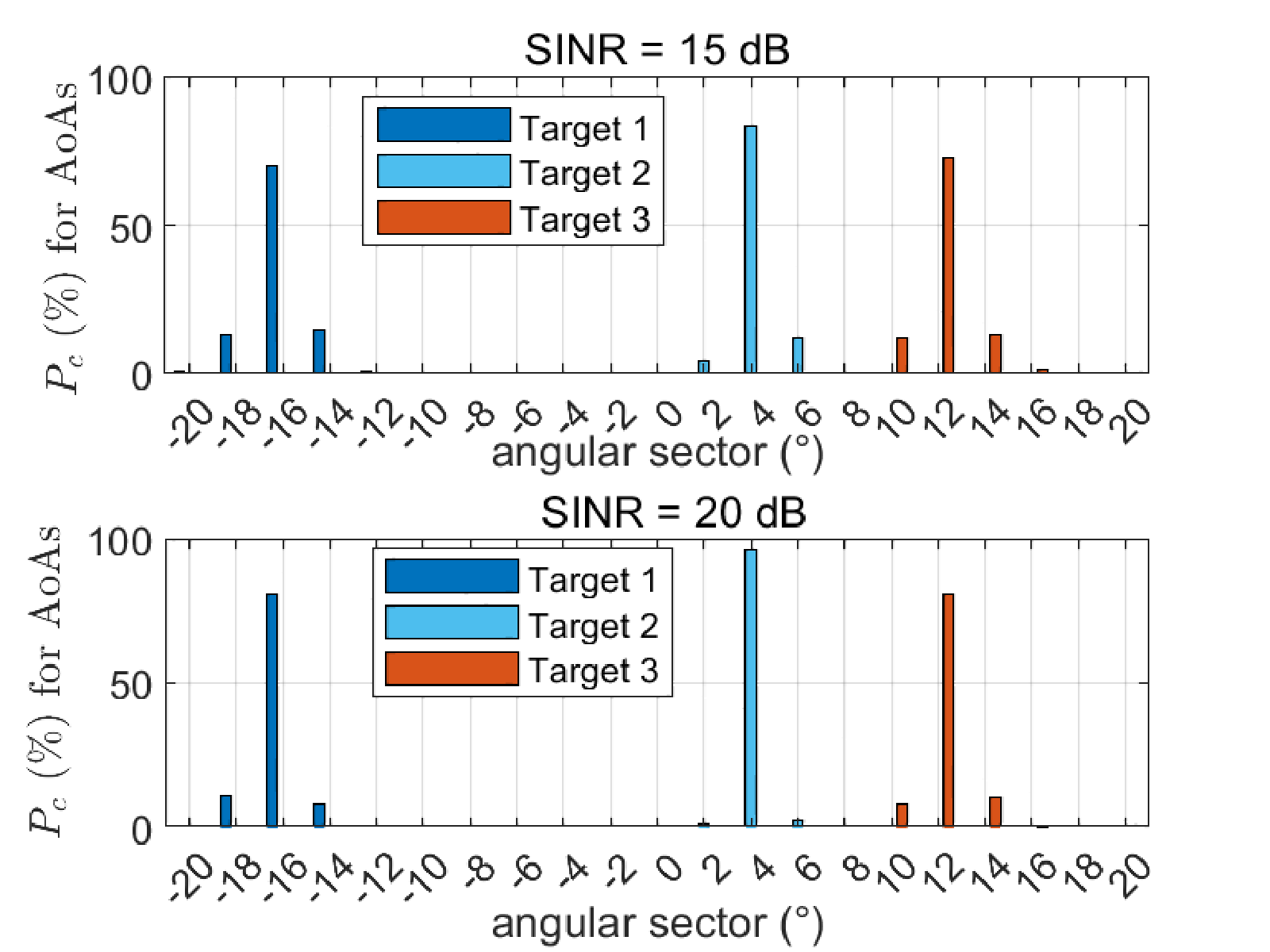}}\\
	\end{center}
	\caption{$P_c$ ($\%$) for angular position resorting to 1000 independent runs (matched AoAs). 
	}
	\label{Pc}
\end{figure}

The analysis, now, proceeds with the evaluation of the classification performance for what concerns the range positions and the AoAs. To this end, in Fig. \ref{snapshot}, we show the classification results obtained in one MC trial for SINR $=15$ dB and $20$ dB, respectively. Note that in Figs. \ref{snapshot} (a) and  \ref{snapshot} (c), class 2 represents the presence of target components, whereas class 1 indicates that the data contain interference only. It can be seen that except for an error of 2$^\circ$ in target AoA estimation at the 13th range for SINR = 15 dB, all the other results perfectly fit the true parameter values, at least for the considered setting. In Fig. \ref{CCP}, a quantitative analysis is conducted  to evaluate the correct classification probability (CCP, $\%$) of target position resorting to 1000 independent runs and leaving the parameters values as specified in Fig. \ref{snapshot}. It is clear that the proposed method demonstrates reliable classification performance since the CCPs for each range achieve no less than 90\% for SINR = 15 dB and increase to almost 100\% when SINR = 20 dB. Moreover, Fig. \ref{Pc} plots the Probability of Classification ($P_c$, $\%$) with respect to the AoAs of the three targets averaged over 1000 trials for SINR values of 15 dB and 20 dB. The figure shows that the percentage of selecting the true AoA values is greater than 70$\%$ when SINR = 15 dB and than 80$\%$ when SINR = 20 dB. 

\begin{figure}[t!]
	\begin{center}
		\subfigure[]{\includegraphics[width=.45\textwidth,height= 6.3 cm]{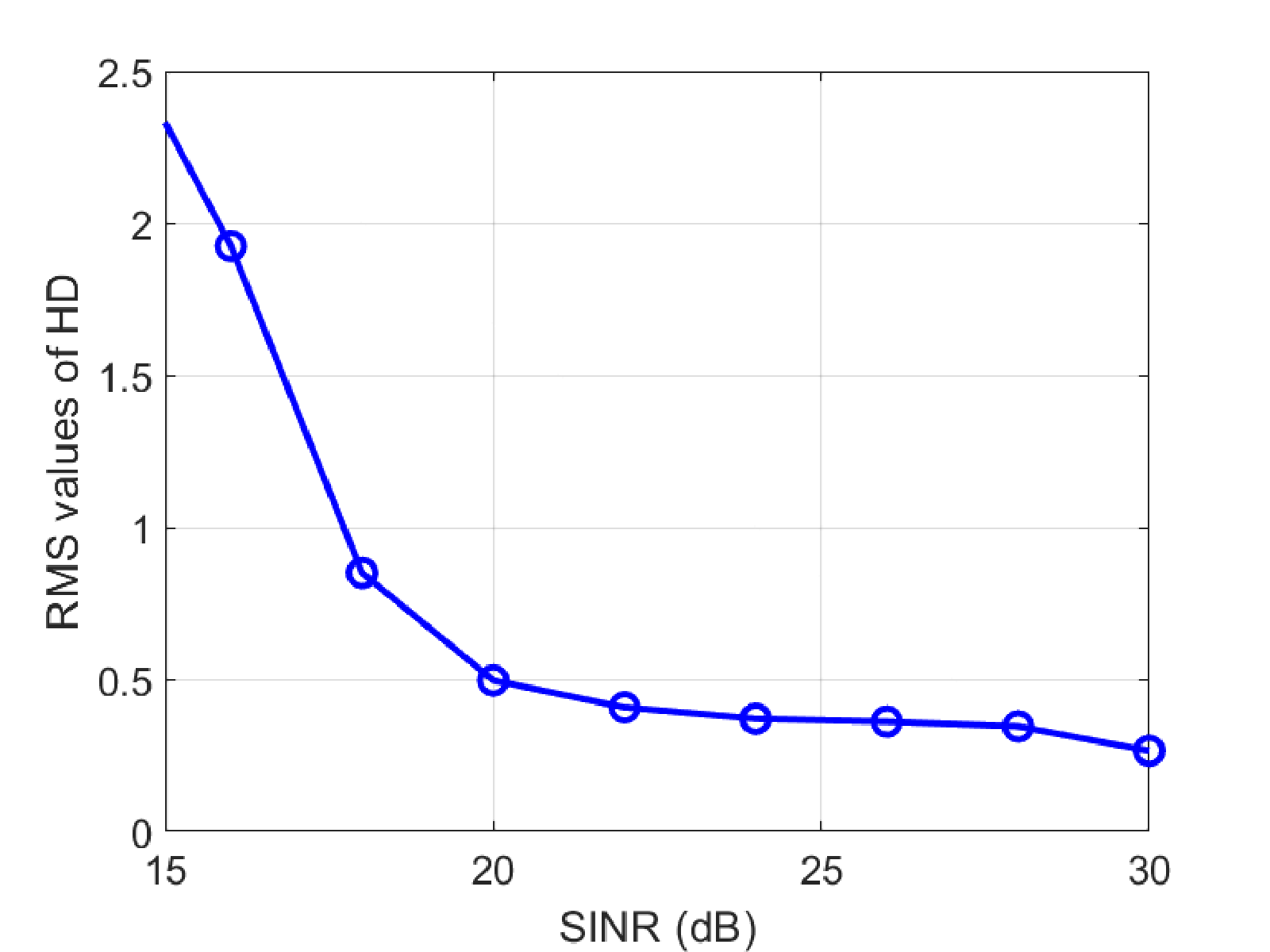}}
		\subfigure[]{\includegraphics[width=.45\textwidth,height= 6.3 cm]{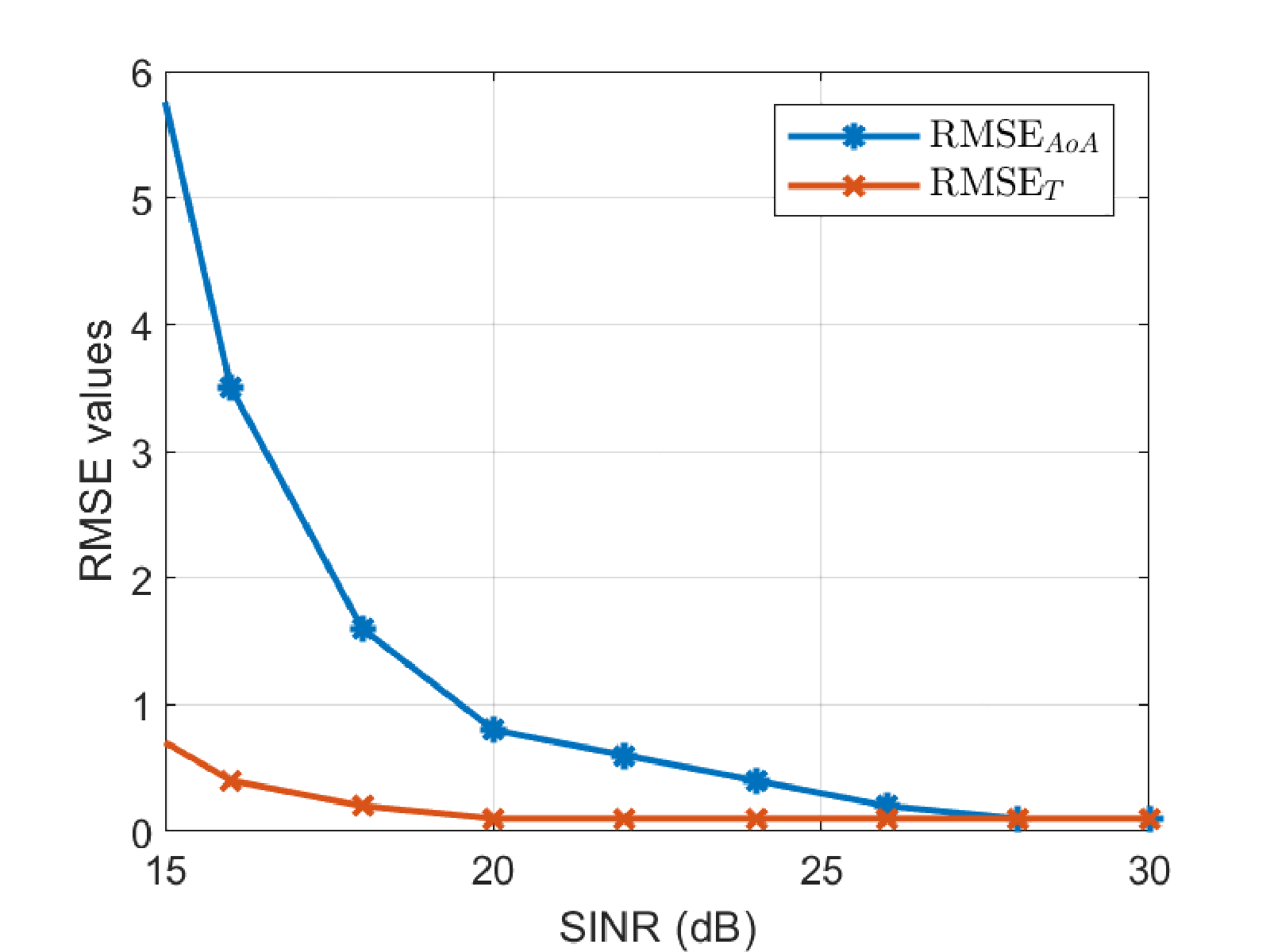}}
	\end{center}
	\caption{ (a) RMS values of HD for range position over 1000 trials; (b) RMSE values for the AoA and number of targets over 1000 trials. (matched AoAs)  
	}
	\label{HD}
\end{figure}

Let us focus on the estimation accuracy for the parameters of interest. Firstly, in Fig. \ref{HD} (a) we plot the Root Mean Square (RMS) values of the Hausdorff Distance (HD) \cite{5744132, 4567674} with respect to  multiple target range positions over 1000 MC trials. More specifically, this distance is defined as $h_d(\cX,\cY) = \max \{ \max_{x \in \cX} \min_{y\in \cY} d(x,y), \max_{y\in \cY} \min_{x\in \cX}  d(x,y)  \}$, where $\cX$ and $\cY$ are the $K$-dimensional sets with nonzero entries indexing the coordinates of the multiple targets in $\widehat{\Omega}_{T}$ and $\Omega_{T}$, respectively. Inspection of the figure reveals that the RMS value decreases as the SINR increases and becomes lower than 0.5 when SINR $\geq$ 20 dB.   In Fig. \ref{HD} (b), we assess the RMS errors (RMSEs) associated with the estimation of the AoAs corresponding to different range bins and the number of targets, which are given by 
\be
RMSE_{AoA} = \sqrt{ \frac{\sum_{j=1}^{N_{MC}} \frac{1}{T} \sum_{t=1}^{T} \min_{\widehat{t} = 1,\ldots,\widehat{T}_{j}}\left(\theta_t - \widehat{\theta}_{\widehat{t}} \right) ^2  } {N_{MC}}   }
\ee
 and 
 \be
 RMSE_{T} = \sqrt{\sum_{j=1}^{N_{MC}} \left(\widehat{T}_{j} - T\right) ^2 /N_{MC}  },
 \ee
  respectively, where $\widehat{\theta}_{\widehat{t}}$  represents the estimates of the $\widehat{t}$th AoA with  $\widehat{T}_{j}$ the estimate of the number of targets at the $j$th MC trial, and $N_{MC}=1000$ denotes the number of independent trials. The results show that the developed procedure attains $RMSE_{AoA}$ and $RMSE_{T}$ less than 1 (degree and number of targets) when SINR $\geq$ 20 dB and SINR $\geq$ 15 dB, respectively. As the SINR increases to 30 dB, both the RMSE errors approach 0.


%
%

\begin{figure}[t!]
	\begin{center}
		\subfigure{\includegraphics[width=.45\textwidth,height= 6.3 cm]{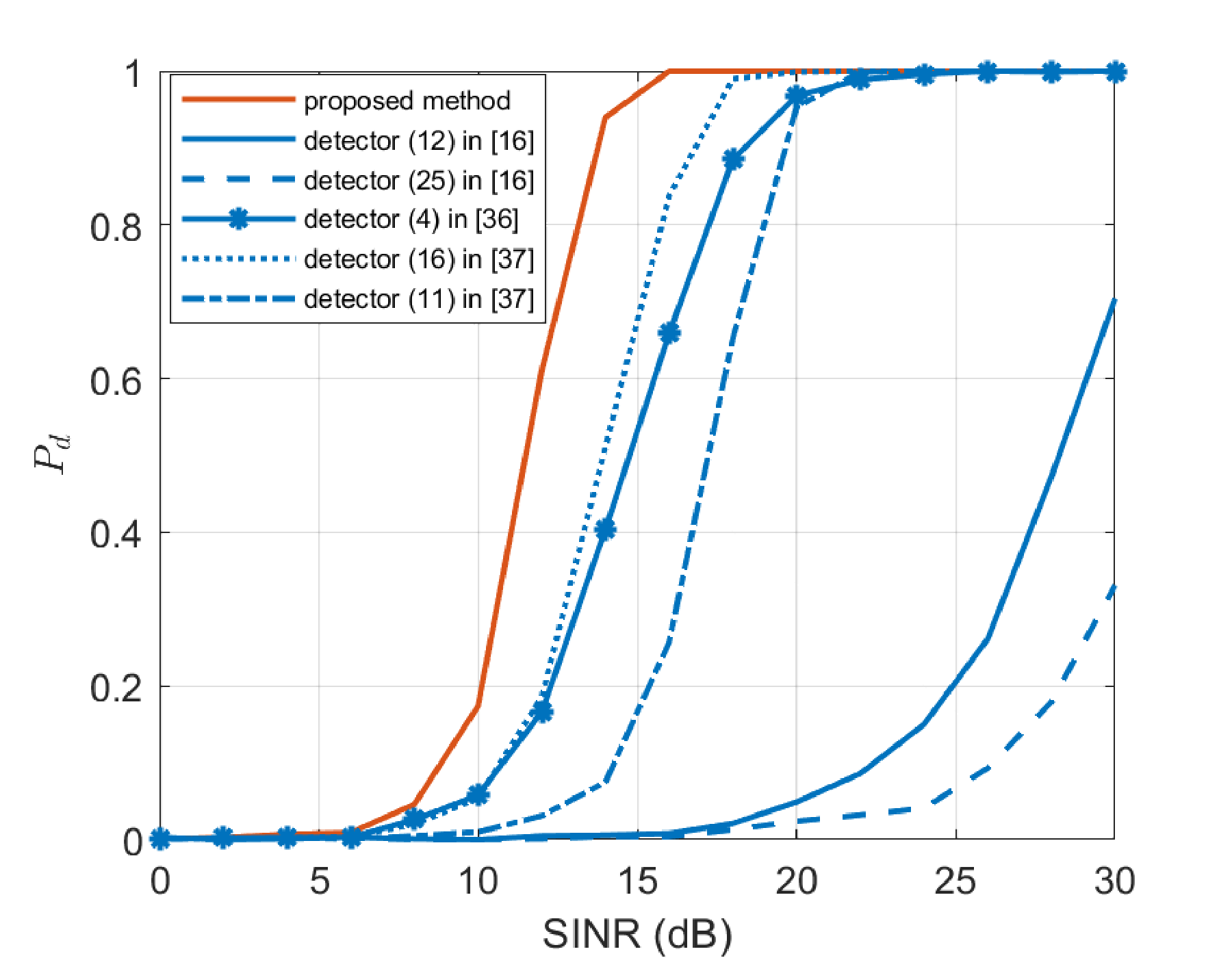}}
	\end{center}
	\caption{$P_d$ versus SINR  for $P_{fa}=10^{-3}$ over 1000 trials (matched AoAs).
	}
	\label{PD}
\end{figure}

In Fig. \ref{PD}, we analyze the detection performance of the proposed detector in comparison with five GLRT-based detectors for homogeneous environment. Specifically, we consider detectors (12) and (25) of \cite{GLRT-based}, detector (4) of \cite{1605248}, and detectors (11) and (16) of \cite{10058041} under $H_{1,1}$ assuming one clutter region. Notice that detectors in \cite{GLRT-based} assume that the target echoes occupy several contiguous range bins, while the other competitors are used to detect multiple point-like targets. Therefore, for detectors (12) and (25) in \cite{GLRT-based}, we consider the range cells from the 4th to the 18th as the primary data window, whereas the remaining cells are used as the secondary data set; moreover, the nominal AoA used in the competitors is equal to 4$^\circ$. Under this configuration, the Probability of Detection ($P_d$) for all the detectors is computed by resorting to 1000 independent MC trials, whereas
	the detection thresholds are computed exploiting 100/$P_{fa}$
	independent MC trials with the Probability of False Alarm $P_{fa}=10^{-3}$. Results show that the proposed detector experiences a superior detection performance
	 compared to other considered decision schemes. Actually, this enhancement arises from the accurate assessment of the multiple targets' parameters, whereas the performance of the considered competitors is affected by a degradation related to the fact that they cannot capitalize all the target energy because of AoA and/or range bin mismatches.  
	
\begin{figure}[t!]
	\begin{center}
		\subfigure[]{\includegraphics[width=.45\textwidth,height= 6.3 cm]{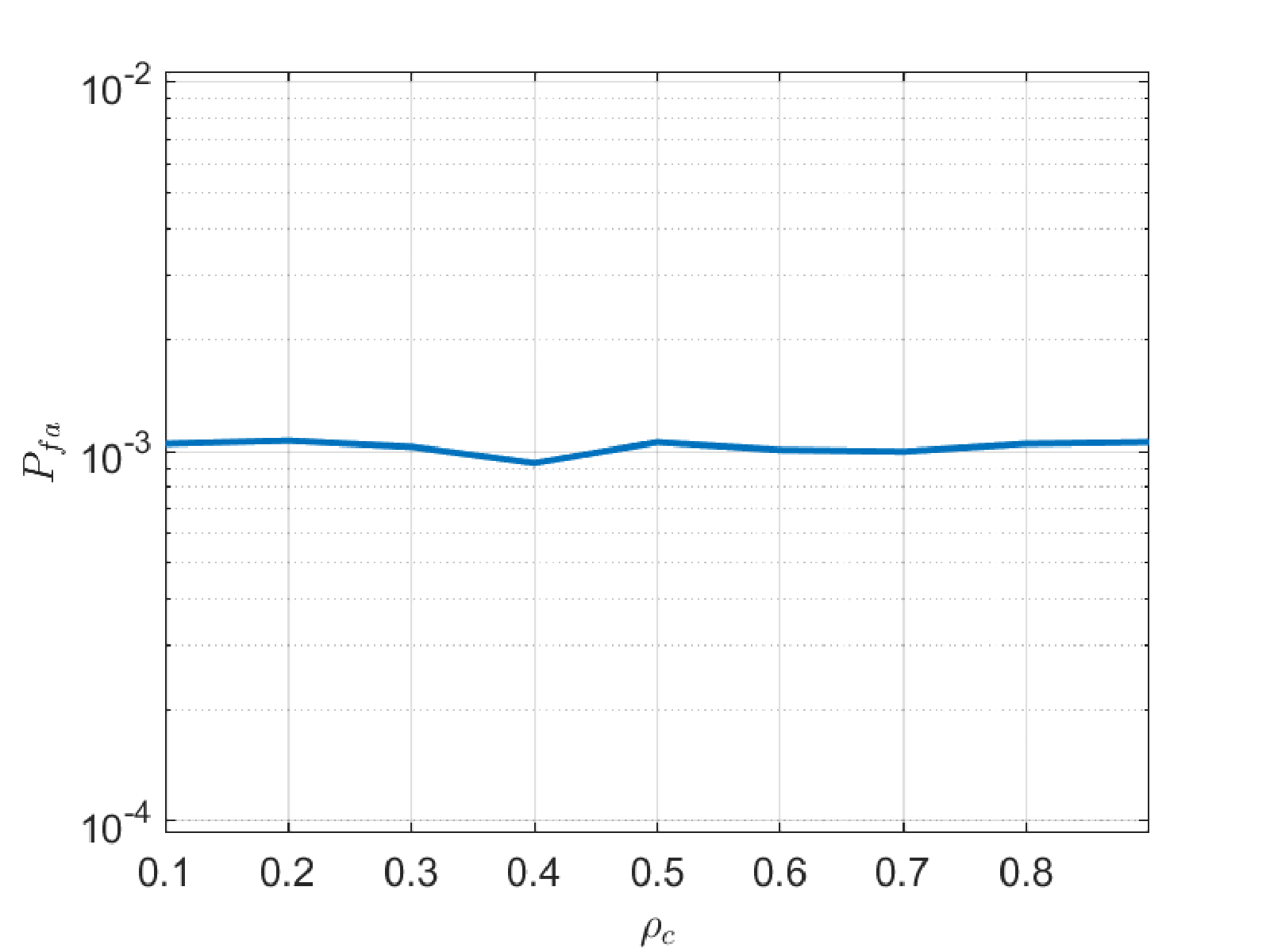}}
		\subfigure[]{\includegraphics[width=.45\textwidth,height= 6.3 cm]{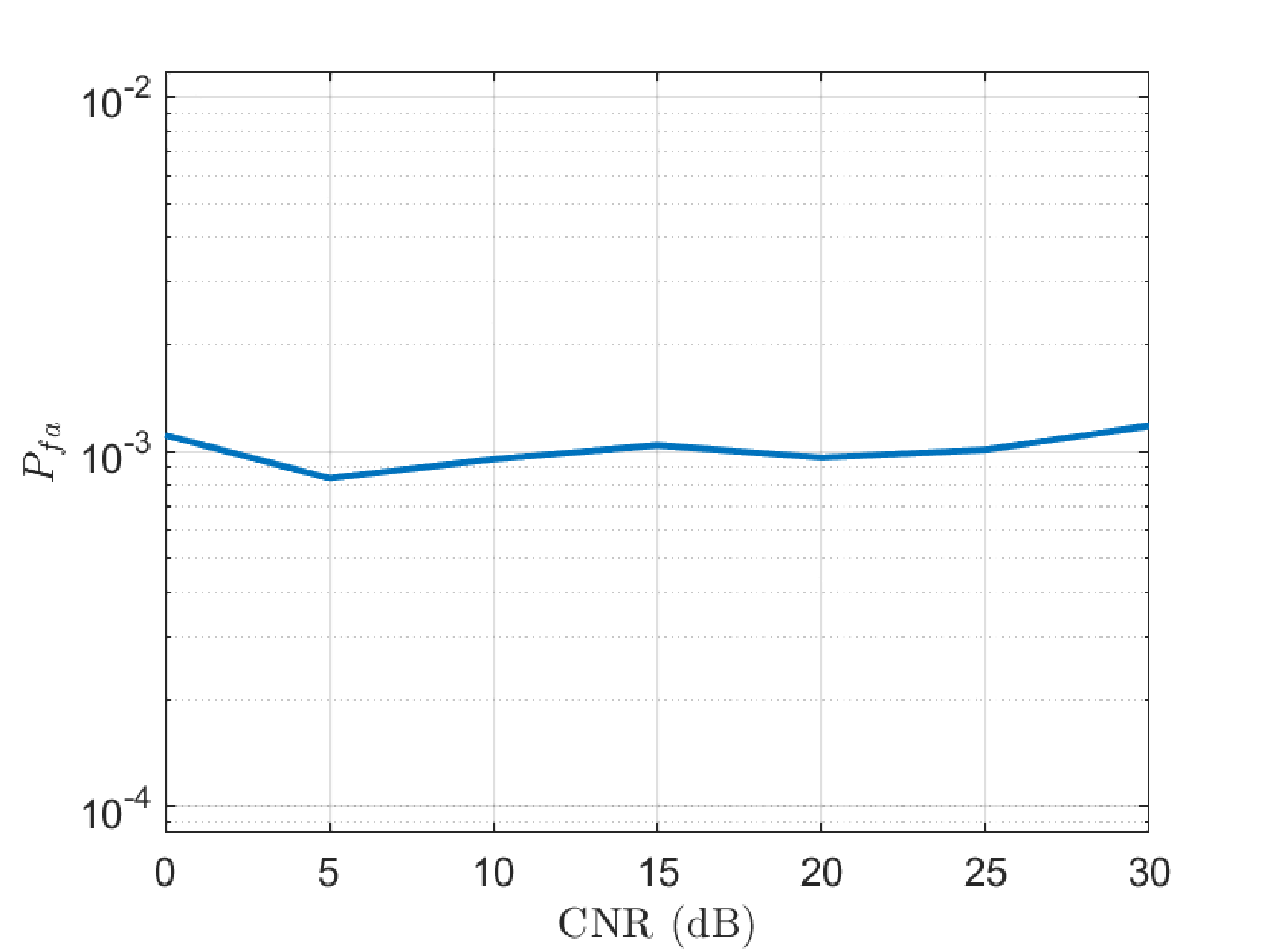}}
	\end{center}
	\caption{ (a) True $P_{fa}$ values versus $\rho_c$; (b) True $P_{fa}$ values versus CNR (matched AoAs).
	}
	\label{CFAR_match}
\end{figure}

	Finally, we focus on the evaluation of the Constant False Alarm Rate (CFAR) behavior of the proposed detector for the considered parameter setting. To this end, in Fig. \ref{CFAR_match}, the $P_{fa}$ is estimated as a function of the CNR and one-lag correlation coefficient $\rho_c$ through 100/$P_{fa}$ MC trials when the threshold is obtained by assuming CNR = 15 dB and $\rho_c = 0.9$. It turns out that the $P_{fa}$ values are practically equal to the nominal value $10^{-3}$ and independent of the clutter parameters at least for the considered ranges of values.


%
%


\begin{figure}[htbp]
	\begin{center}
		{\includegraphics[width=.45\textwidth,height= 6.3 cm]{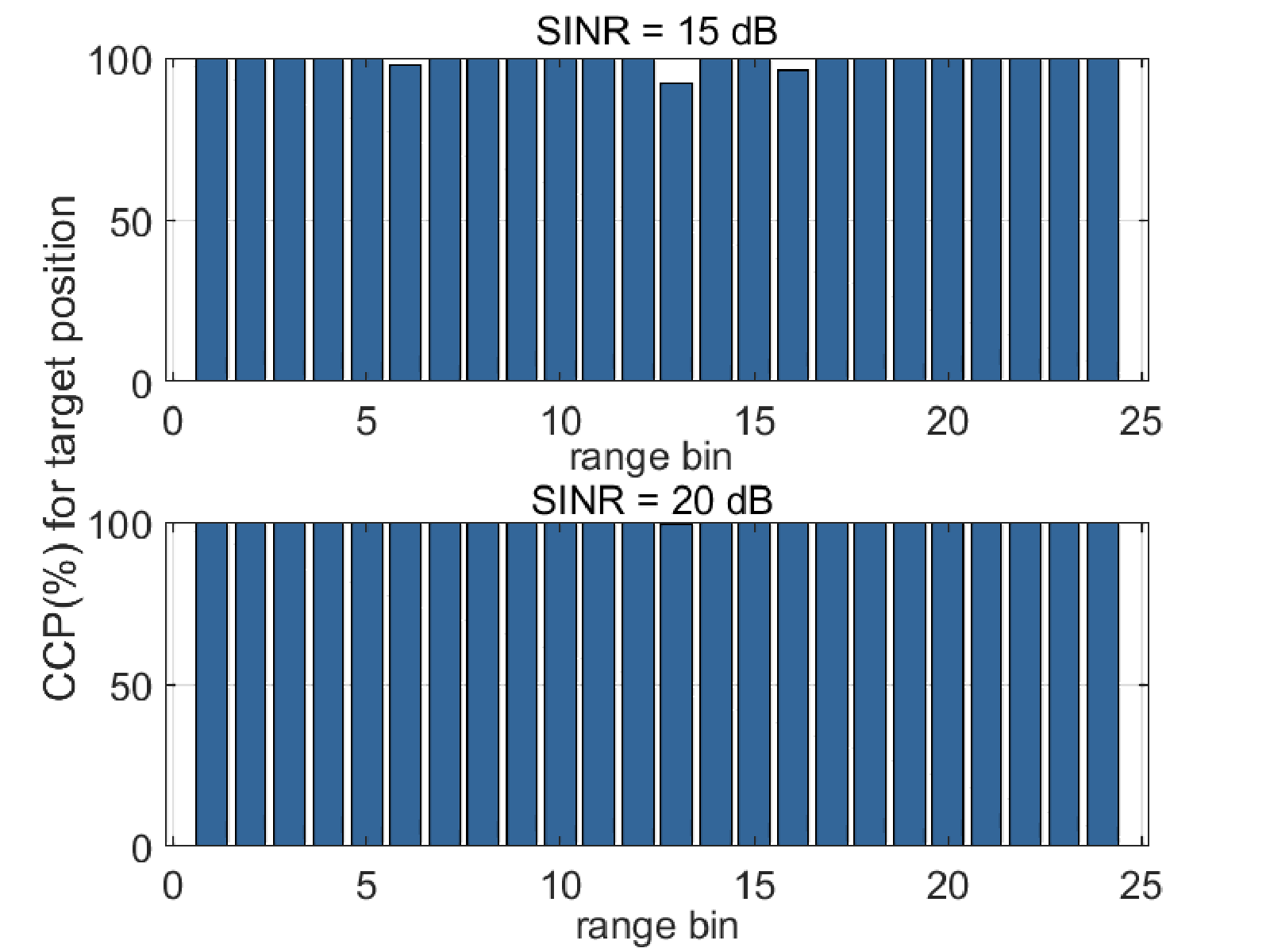}}
	\end{center}
	\caption{CCP ($\%$) for target position resorting to 1000 independent runs (mismatched AoAs). 
	}
	\label{CCP_mismatch}
\end{figure}

\begin{figure}[htbp]
	\begin{center}
		{\includegraphics[width=.45\textwidth,height= 6.3 cm]{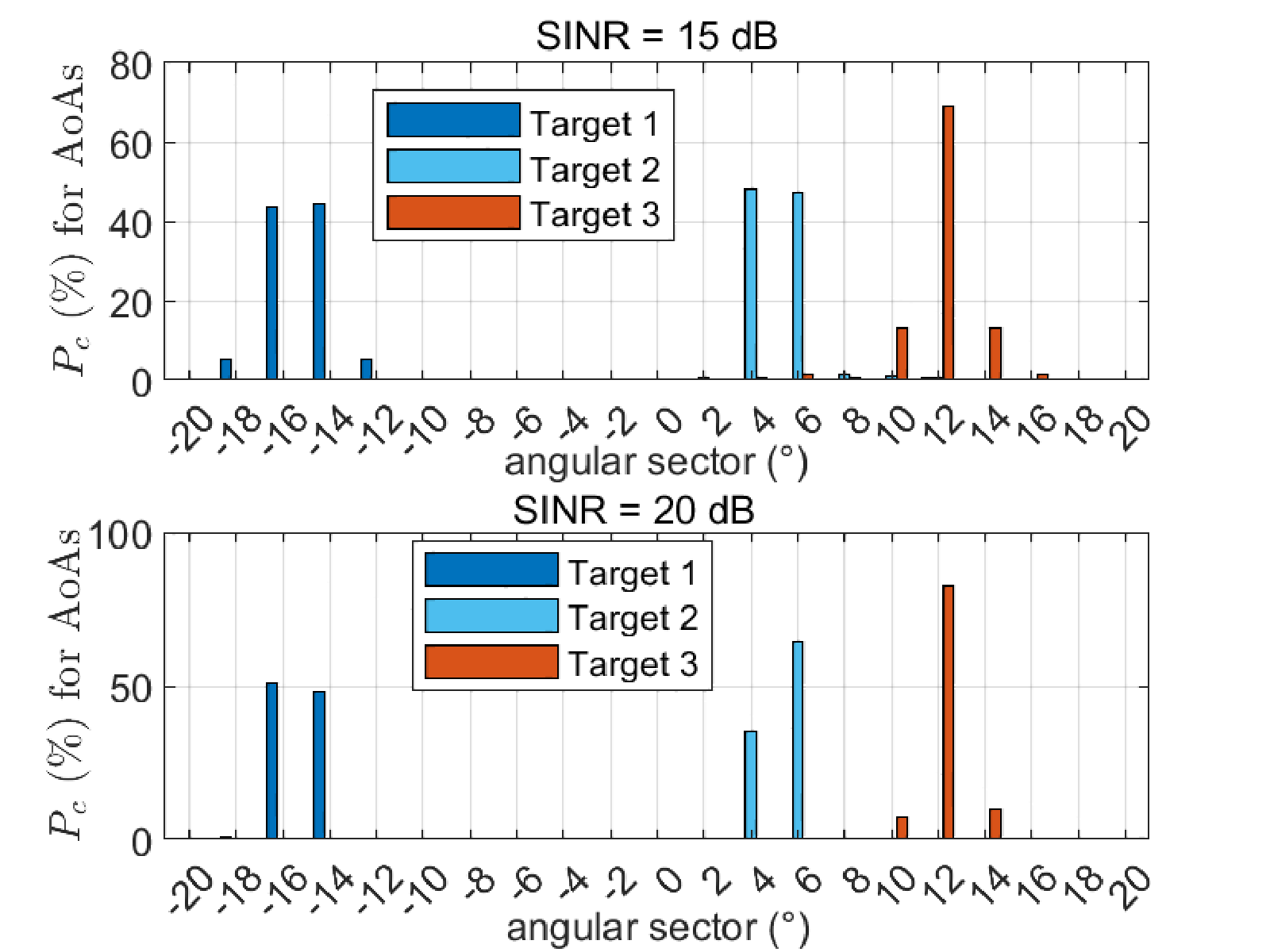}}
	\end{center}
	\caption{$P_c$ ($\%$) for angular position resorting to 1000 independent runs (mismatched AoAs). 
	}
	\label{Pc_mismatch}
\end{figure}

\subsection{Mismatched Target AoAs}

\begin{figure}[htbp]
	\begin{center}
		\subfigure[]{\includegraphics[width=.45\textwidth,height= 6.3 cm]{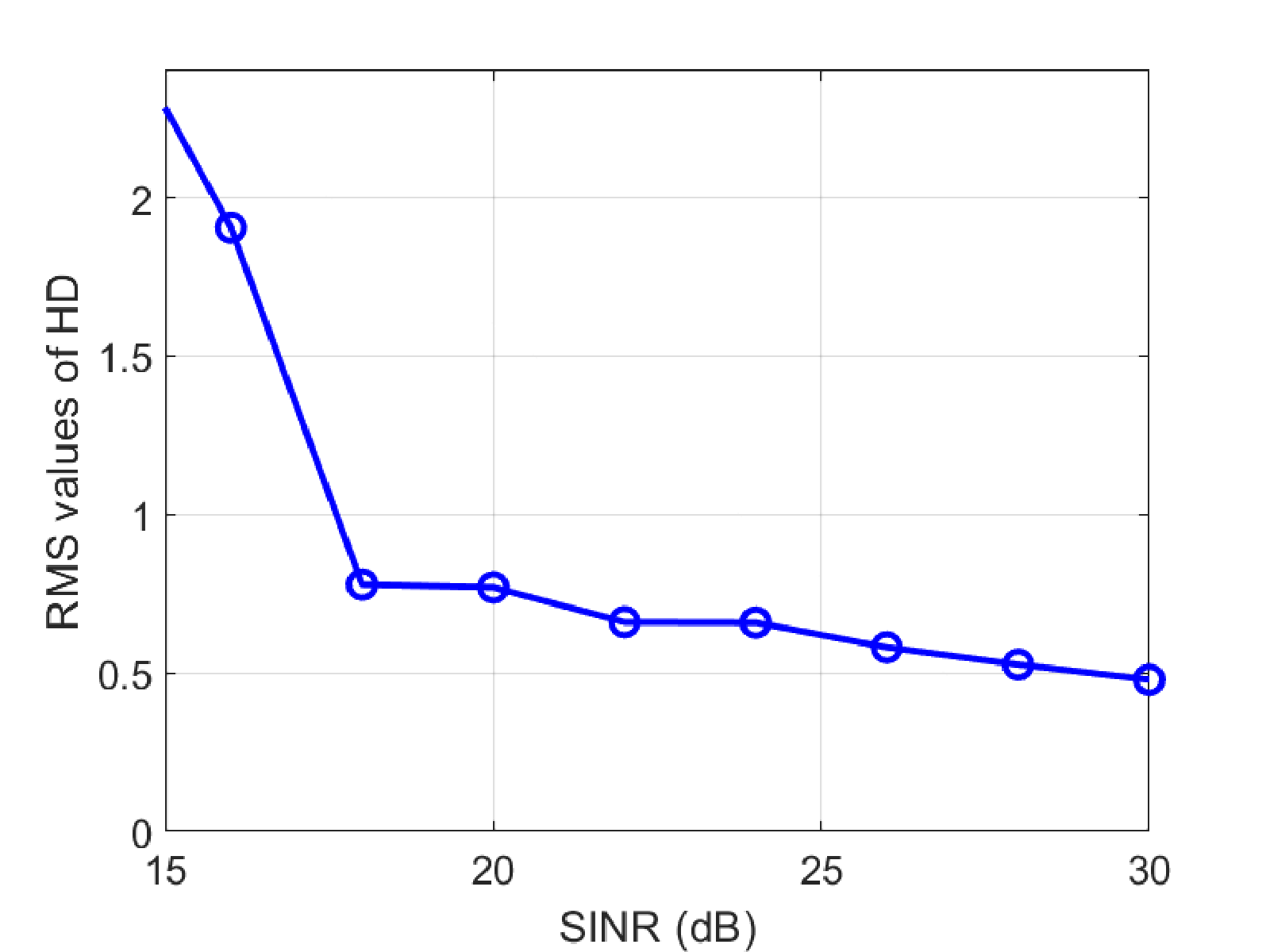}}
		\subfigure[]{\includegraphics[width=.45\textwidth,height= 6.3 cm]{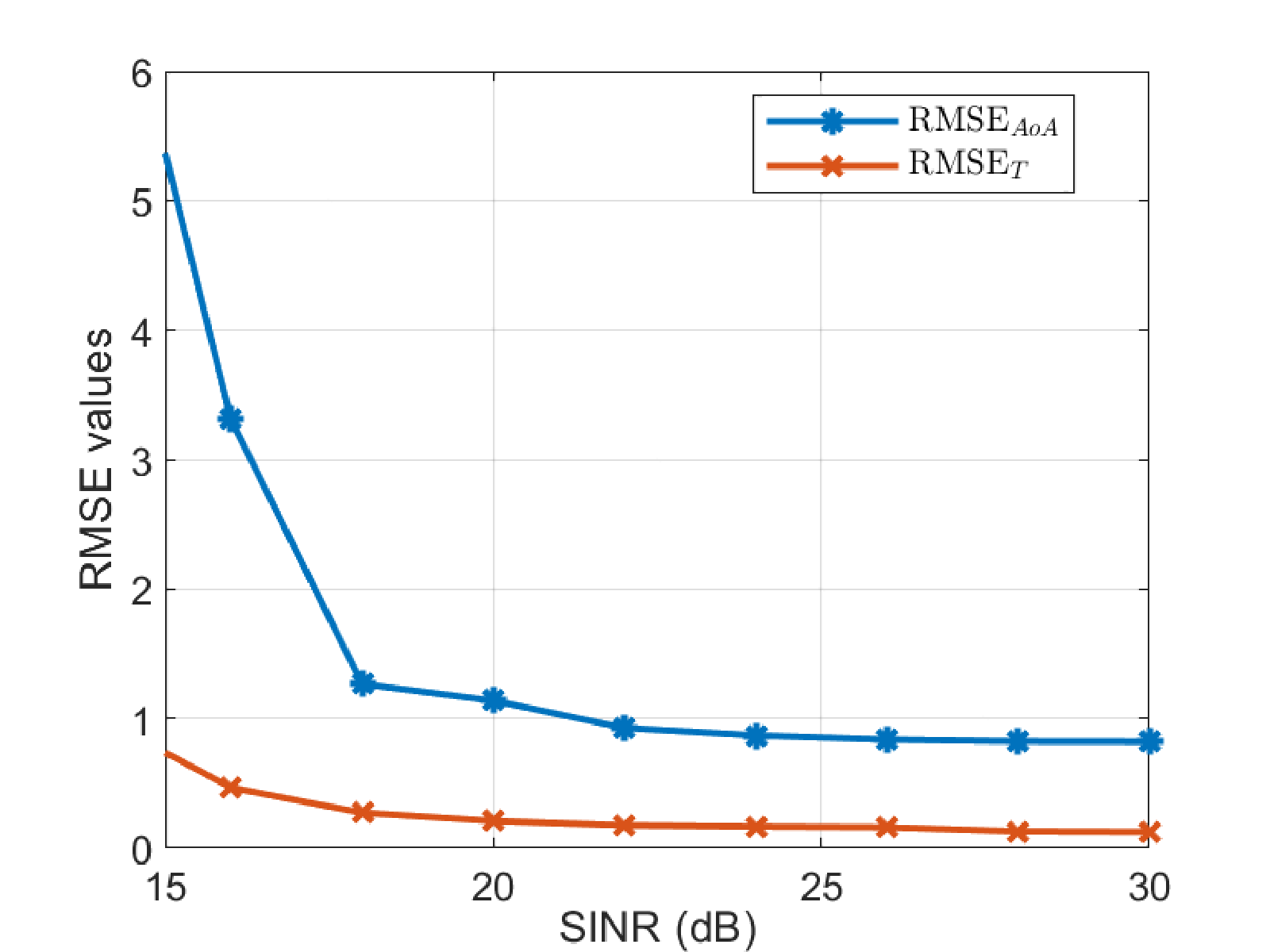}}
	\end{center}
	\caption{ (a) RMS values of HD for range position over 1000 trials; (b) RMSE values for the AoA and number of targets over 1000 trials. (mismatched AoAs)  
	}
	\label{HD_mismatched}
\end{figure}

\begin{figure}[t!]
	\begin{center}
		\subfigure{\includegraphics[width=.45\textwidth,height= 6.3 cm]{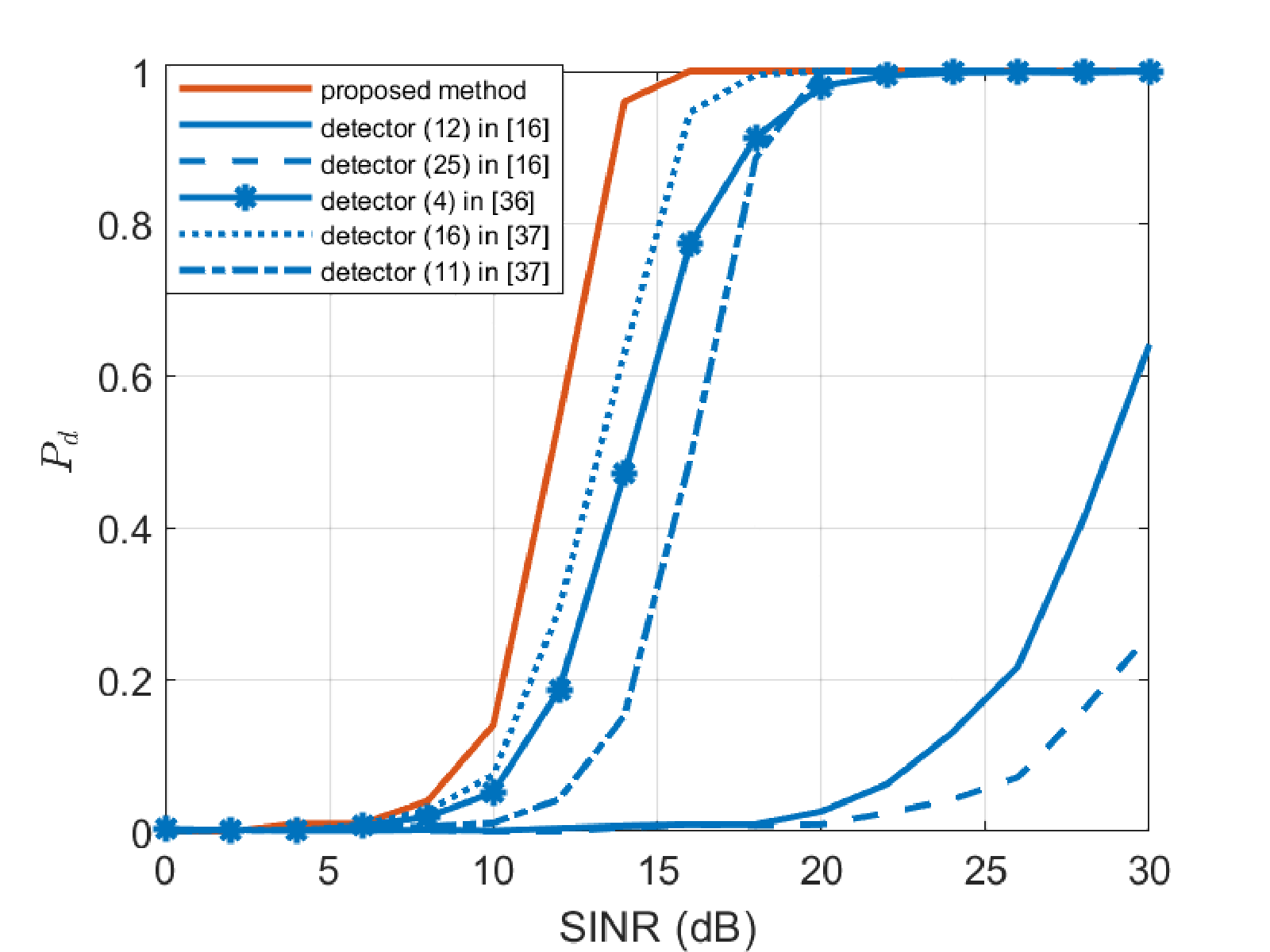}}
	\end{center}
	\caption{$P_d$ versus SINR  for $P_{fa}=10^{-3}$ over 1000 trials (mismatched AoAs). 
	}
	\label{PD_mismatch}
\end{figure}

In this subsection, we assess the performance of the newly proposed method  in the presence of AoA mismatches. To this end, we analyze a scenario where two out of the three targets in Table \ref{T1} are positioned between the grid points of $\Omega_{\theta}$, namely the true AoAs are -15$^\circ$, 5$^\circ$, and 12$^\circ$. Figs. \ref{CCP_mismatch}-\ref{Pc_mismatch} present the CCP ($\%$) for target position and the $P_c$ ($\%$) for angular position, respectively, for SINR values of 15 dB and 20 dB, based on 1000 independent MC trials. Fig. \ref{CCP_mismatch} shows that the developed scheme for this scenario exhibits an excellent performance for the classification of the range bins. As for the selected AoA from the grid, the estimates mostly converge on the grid points that are closest to the true AoAs. In Fig. \ref{HD_mismatched}, we plot the same metrics as in Fig. \ref{HD} to evaluate the estimation quality for the range position, angular position and number of targets. It can be seen that the RMS-based metrics approach very low values for relatively high SINR values. Finally, we investigate the detection performance of the proposed detector in comparison with previously mentioned couterparts. The norminal pointing direction for the competitors is set to 5$^{\circ}$. In Fig. \ref{PD_mismatch}, the $P_d$ curves demonstrate the performance advantage also in the presence of AoA mismatches.

\section{conclusion}
In this paper, we have proposed a new adaptive decision scheme to address
the detection
of an unknown number of point-like targets deployed in the range-azimuth map. To this end, a binary hypothesis test is first
formulated by using a modification of the LVM
that in this context accounts for the probability of occurrence of a
target within each range cell and  angular sector. At the design stage,
the EM algorithm
is exploited to devise an iterative estimation procedure capable of
providing reliable estimates for the unknown
parameters and also serving for the design of an LRT-based
detector. Numerical examples obtained through synthetic data
corroborate the effectiveness of the proposed architecture
in terms of detection and estimation performance as well
as CFAR  behavior (at least for the considered ranges of parameters). Remarkably, the proposed LVM-based
approach exhibits a significant advantage with respect to the
considered competitors.  
      
     Possible extensions of this research line could concern  robust detection of multiple point-like targets in heterogeneous scenarios where clutter edges  or  non-Gaussian interference are present.

\bibliographystyle{IEEEtran}
\bibliography{group_bib}

\begin{thebibliography}{10}
\providecommand{\url}[1]{#1}
\csname url@samestyle\endcsname
\providecommand{\newblock}{\relax}
\providecommand{\bibinfo}[2]{#2}
\providecommand{\BIBentrySTDinterwordspacing}{\spaceskip=0pt\relax}
\providecommand{\BIBentryALTinterwordstretchfactor}{4}
\providecommand{\BIBentryALTinterwordspacing}{\spaceskip=\fontdimen2\font plus
\BIBentryALTinterwordstretchfactor\fontdimen3\font minus
  \fontdimen4\font\relax}
\providecommand{\BIBforeignlanguage}[2]{{%
\expandafter\ifx\csname l@#1\endcsname\relax
\typeout{** WARNING: IEEEtran.bst: No hyphenation pattern has been}%
\typeout{** loaded for the language `#1'. Using the pattern for}%
\typeout{** the default language instead.}%
\else
\language=\csname l@#1\endcsname
\fi
#2}}
\providecommand{\BIBdecl}{\relax}
\BIBdecl

\bibitem{5438467}
A.~De~Maio, S.~De~Nicola, A.~Farina, and S.~Iommelli, ``A robust adaptive
  detector for steering phase uncertainties,'' in \emph{2009 International
  Radar Conference "Surveillance for a Safer World" (RADAR 2009)}, 2009, pp.
  1--6.

\bibitem{2010Adaptive}
A.~De~Maio, S.~D. Nicola, A.~Farina, and S.~Iommelli, ``Adaptive detection of a
  signal with angle uncertainty,'' \emph{IET Radar, Sonar and Navigation},
  vol.~4, no.~4, pp. 537--547, 2010.

\bibitem{4104190}
E.~Kelly, ``An adaptive detection algorithm,'' \emph{IEEE Transactions on
  Aerospace and Electronic Systems}, vol. AES-22, no.~2, pp. 115--127, 1986.

\bibitem{135446}
F.~Robey, D.~Fuhrmann, E.~Kelly, and R.~Nitzberg, ``A {CFAR} adaptive matched
  filter detector,'' \emph{IEEE Transactions on Aerospace and Electronic
  Systems}, vol.~28, no.~1, pp. 208--216, 1992.

\bibitem{599116}
L.~Scharf and L.~McWhorter, ``Adaptive matched subspace detectors and adaptive
  coherence estimators,'' in \emph{Conference Record of The Thirtieth Asilomar
  Conference on Signals, Systems and Computers}, 1996, pp. 1114--1117 vol.2.

\bibitem{4102349}
D.~M. Boroson, ``Sample size considerations for adaptive arrays,'' \emph{IEEE
  Transactions on Aerospace and Electronic Systems}, vol. AES-16, no.~4, pp.
  446--451, 1980.

\bibitem{18674}
E.~Kelly, ``Performance of an adaptive detection algorithm; rejection of
  unwanted signals,'' \emph{IEEE Transactions on Aerospace and Electronic
  Systems}, vol.~25, no.~2, pp. 122--133, 1989.

\bibitem{7303927}
J.~Liu, W.~Liu, B.~Chen, H.~Liu, and H.~Li, ``Detection probability of a {CFAR}
  matched filter with signal steering vector errors,'' \emph{IEEE Signal
  Processing Letters}, vol.~22, no.~12, pp. 2474--2478, 2015.

\bibitem{301849}
L.~Scharf and B.~Friedlander, ``Matched subspace detectors,'' \emph{IEEE
  Transactions on Signal Processing}, vol.~42, no.~8, pp. 2146--2157, 1994.

\bibitem{381913}
R.~Raghavan, H.~Qiu, and D.~McLaughlin, ``{CFAR} detection in clutter with
  unknown correlation properties,'' \emph{IEEE Transactions on Aerospace and
  Electronic Systems}, vol.~31, no.~2, pp. 647--657, 1995.

\bibitem{492544}
K.~Burgess and B.~Van~Veen, ``Subspace-based adaptive generalized likelihood
  ratio detection,'' \emph{IEEE Transactions on Signal Processing}, vol.~44,
  no.~4, pp. 912--927, 1996.

\bibitem{890324}
S.~Kraut, L.~Scharf, and L.~McWhorter, ``Adaptive subspace detectors,''
  \emph{IEEE Transactions on Signal Processing}, vol.~49, no.~1, pp. 1--16,
  2001.

\bibitem{VanTrees2002}
H.~L.~V. Trees, \emph{Detection, Estimation, and Modulation Theory, Part IV:
  Optimum Array Processing}, ser. Detection, Estimation, and Modulation
  Theory.\hskip 1em plus 0.5em minus 0.4em\relax Wiley-Interscience, 2002,
  vol.~4.

\bibitem{4014367}
O.~Besson, L.~L. Scharf, and S.~Kraut, ``Adaptive detection of a signal known
  only to lie on a line in a known subspace, when primary and secondary data
  are partially homogeneous,'' \emph{IEEE Transactions on Signal Processing},
  vol.~54, no.~12, pp. 4698--4705, 2006.

\bibitem{1542473}
O.~Besson, L.~Scharf, and F.~Vincent, ``Matched direction detectors and
  estimators for array processing with subspace steering vector
  uncertainties,'' \emph{IEEE Transactions on Signal Processing}, vol.~53,
  no.~12, pp. 4453--4463, 2005.

\bibitem{GLRT-based}
E.~Conte, A.~De~Maio, and G.~Ricci, ``{GLRT-based adaptive detection algorithms
  for range-spread targets},'' \emph{IEEE Transactions on Signal Processing},
  vol.~49, no.~7, pp. 1336--1348, July 2001.

\bibitem{1145714}
E.~Conte, A.~De~Maio, and C.~Galdi, ``{CFAR} detection of multidimensional
  signals: an invariant approach,'' \emph{IEEE Transactions on Signal
  Processing}, vol.~51, no.~1, pp. 142--151, 2003.

\bibitem{6576130}
W.~Liu, W.~Xie, and Y.~Wang, ``{Rao} and {Wald} tests for distributed targets
  detection with unknown signal steering,'' \emph{IEEE Signal Processing
  Letters}, vol.~20, no.~11, pp. 1086--1089, 2013.

\bibitem{1396421}
Y.~Jin and B.~Friedlander, ``A {CFAR} adaptive subspace detector for
  second-order gaussian signals,'' \emph{IEEE Transactions on Signal
  Processing}, vol.~53, no.~3, pp. 871--884, 2005.

\bibitem{4531367}
F.~Bandiera, D.~Orlando, and G.~Ricci, ``A subspace-based adaptive sidelobe
  blanker,'' \emph{IEEE Transactions on Signal Processing}, vol.~56, no.~9, pp.
  4141--4151, 2008.

\bibitem{5978230}
J.~Liu, Z.-J. Zhang, Y.~Yang, and H.~Liu, ``A {CFAR} adaptive subspace detector
  for first-order or second-order gaussian signals based on a single
  observation,'' \emph{IEEE Transactions on Signal Processing}, vol.~59,
  no.~11, pp. 5126--5140, 2011.

\bibitem{6879481}
S.~Lei, Z.~Zhao, Z.~Nie, and Q.-H. Liu, ``A {CFAR} adaptive subspace detector
  based on a single observation in system-dependent clutter background,''
  \emph{IEEE Transactions on Signal Processing}, vol.~62, no.~20, pp.
  5260--5269, 2014.

\bibitem{7938714}
H.~Li, Y.~Jiang, J.~Fang, and M.~Rangaswamy, ``Adaptive subspace signal
  detection with uncertain partial prior knowledge,'' \emph{IEEE Transactions
  on Signal Processing}, vol.~65, no.~16, pp. 4394--4405, 2017.

\bibitem{1561887}
A.~De~Maio, ``Robust adaptive radar detection in the presence of steering
  vector mismatches,'' \emph{IEEE Transactions on Aerospace and Electronic
  Systems}, vol.~41, no.~4, pp. 1322--1337, 2005.

\bibitem{4203039}
F.~Bandiera, A.~De~Maio, and G.~Ricci, ``Adaptive {CFAR} radar detection with
  conic rejection,'' \emph{IEEE Transactions on Signal Processing}, vol.~55,
  no.~6, pp. 2533--2541, 2007.

\bibitem{4895239}
F.~Bandiera, D.~Orlando, and G.~Ricci, ``{CFAR} detection strategies for
  distributed targets under conic constraints,'' \emph{IEEE Transactions on
  Signal Processing}, vol.~57, no.~9, pp. 3305--3316, 2009.

\bibitem{8716559}
L.~Shen, Z.~Liu, Y.~Xu, Y.~Bai, and T.~Zhao, ``Robust polarimetric adaptive
  detector against target steering matrix mismatch,'' \emph{IEEE Transactions
  on Aerospace and Electronic Systems}, vol.~56, no.~1, pp. 442--455, 2020.

\bibitem{839972}
C.~Richmond, ``Performance of the adaptive sidelobe blanker detection algorithm
  in homogeneous environments,'' \emph{IEEE Transactions on Signal Processing},
  vol.~48, no.~5, pp. 1235--1247, 2000.

\bibitem{4531361}
F.~Bandiera, O.~Besson, D.~Orlando, and G.~Ricci, ``An improved adaptive
  sidelobe blanker,'' \emph{IEEE Transactions on Signal Processing}, vol.~56,
  no.~9, pp. 4152--4161, 2008.

\bibitem{5165249}
F.~Bandiera, D.~Orlando, and G.~Ricci, ``One- and two-stage tunable
  receivers*,'' \emph{IEEE Transactions on Signal Processing}, vol.~57, no.~8,
  pp. 3264--3273, 2009.

\bibitem{2011Performance}
C.~Hao, B.~Liu, and L.~Cai, ``Performance analysis of a two-stage {Rao}
  detector,'' \emph{Signal Processing}, vol.~91, no.~8, pp. 2141--2146, 2011.

\bibitem{8781902}
L.~Yan, P.~Addabbo, C.~Hao, D.~Orlando, and A.~Farina, ``New {ECCM} techniques
  against noiselike and/or coherent interferers,'' \emph{IEEE Transactions on
  Aerospace and Electronic Systems}, vol.~56, no.~2, pp. 1172--1188, 2020.

\bibitem{5617289}
P.~Stoica, P.~Babu, and J.~Li, ``Spice: A sparse covariance-based estimation
  method for array processing,'' \emph{IEEE Transactions on Signal Processing},
  vol.~59, no.~2, pp. 629--638, 2011.

\bibitem{9076078}
L.~Yan, P.~Addabbo, Y.~Zhang, C.~Hao, J.~Liu, J.~Li, and D.~Orlando, ``A sparse
  learning approach to the detection of multiple noise-like jammers,''
  \emph{IEEE Transactions on Aerospace and Electronic Systems}, vol.~56, no.~6,
  pp. 4367--4383, 2020.

\bibitem{9040449}
S.~Han, L.~Pallotta, X.~Huang, G.~Giunta, and D.~Orlando, ``A sparse learning
  approach to the design of radar tunable architectures with enhanced
  selectivity properties,'' \emph{IEEE Transactions on Aerospace and Electronic
  Systems}, vol.~56, no.~5, pp. 3840--3853, 2020.

\bibitem{1605248}
F.~Bandiera, D.~Orlando, and G.~Ricci, ``{CFAR} detection of extended and
  multiple point-like targets without assignment of secondary data,''
  \emph{IEEE Signal Processing Letters}, vol.~13, no.~4, pp. 240--243, 2006.

\bibitem{10058041}
L.~Yan, S.~Han, C.~Hao, D.~Orlando, and G.~Ricci, ``Innovative cognitive
  approaches for joint radar clutter classification and multiple target
  detection in heterogeneous environments,'' \emph{IEEE Transactions on Signal
  Processing}, vol.~71, pp. 1010--1022, 2023.

\bibitem{9048459}
M.~Xu, B.~Chen, J.~Liu, L.~Tian, and P.~Ma, ``Radar {HRRP} target recognition
  based on variational auto-encoder with learnable prior,'' in \emph{2019 6th
  Asia-Pacific Conference on Synthetic Aperture Radar (APSAR)}, 2019, pp. 1--5.

\bibitem{10465106}
I.~W.~G. Da~Silva, D.~P.~M. Osorio, and M.~Juntti, ``Multi-static {ISAC} in
  cell-free massive {MIMO}: Precoder design and privacy assessment,'' in
  \emph{2023 IEEE Globecom Workshops (GC Wkshps)}, 2023, pp. 461--466.

\bibitem{murphy2012machine}
K.~Murphy, \emph{{Machine Learning: A Probabilistic Perspective}}, ser.
  Adaptive Computation and Machine Learning series.\hskip 1em plus 0.5em minus
  0.4em\relax MIT Press, 2012.

\bibitem{10645068}
L.~Yan, P.~Addabbo, N.~Fiscante, C.~Clemente, C.~Hao, G.~Giunta, and
  D.~Orlando, ``{EM}-based algorithm for unsupervised clustering of
  measurements from a radar sensor network,'' \emph{IEEE Transactions on
  Aerospace and Electronic Systems}, pp. 1--15, 2024.

\bibitem{9444189}
S.~Yan, F.~Lotfi, S.~Chen, C.~Hao, and D.~Orlando, ``Innovative two-stage radar
  detection architectures in adverse scenarios using two training data sets,''
  \emph{IEEE Signal Processing Letters}, vol.~28, pp. 1165--1169, 2021.

\bibitem{2021Learning}
P.~Addabbo, S.~Han, D.~Orlando, and G.~Ricci, ``Learning strategies for radar
  clutter classification,'' \emph{IEEE Transactions on Signal Processing},
  2021.

\bibitem{mirsky1959trace}
L.~Mirsky, ``{On the Trace of Matrix Products},'' \emph{Mathematische
  Nachrichten}, vol.~20, pp. 171--174, 1959.

\bibitem{lutkepohl1997handbook}
H.~L{\"u}tkepohl, \emph{{Handbook of Matrices}}.\hskip 1em plus 0.5em minus
  0.4em\relax Wiley, 1997.

\bibitem{1336827}
A.~De~Maio, ``A new derivation of the adaptive matched filter,'' \emph{IEEE
  Signal Processing Letters}, vol.~11, no.~10, pp. 792--793, 2004.

\bibitem{8496824}
J.~Liu, W.~Liu, Y.~Gao, S.~Zhou, and X.-G. Xia, ``Persymmetric adaptive
  detection of subspace signals: Algorithms and performance analysis,''
  \emph{IEEE Transactions on Signal Processing}, vol.~66, no.~23, pp.
  6124--6136, 2018.

\bibitem{7376203}
C.~Hao, S.~Gazor, G.~Foglia, B.~Liu, and C.~Hou, ``Persymmetric adaptive
  detection and range estimation of a small target,'' \emph{IEEE Transactions
  on Aerospace and Electronic Systems}, vol.~51, no.~4, pp. 2590--2604, 2015.

\bibitem{5744132}
B.~Ristic, B.-N. Vo, D.~Clark, and B.-T. Vo, ``A metric for performance
  evaluation of multi-target tracking algorithms,'' \emph{IEEE Transactions on
  Signal Processing}, vol.~59, no.~7, pp. 3452--3457, 2011.

\bibitem{4567674}
D.~Schuhmacher, B.-T. Vo, and B.-N. Vo, ``A consistent metric for performance
  evaluation of multi-object filters,'' \emph{IEEE Transactions on Signal
  Processing}, vol.~56, no.~8, pp. 3447--3457, 2008.

\end{thebibliography}

\end{document}